# Little Black Holes as Dark Matter Candidates
# with
# Feasible Cosmic and Terrestrial Interactions

**Mario Rabinowitz**
Armor Research
715 Lakemead Way
Redwood City, CA 94062-3922
Mario715@earthlink.net

**Abstract**

In addition to contributing to the accelerated expansion of the Universe, little black holes (LBH) can exhibit strong interactions with large-scale manifestations in interacting with each other, and with large macroscopic bodies such as stars, neutron stars, and planets. A range of proposals are reviewed in which both free and bound LBH are considered to be either a small component or the dominant constituent of dark matter/dark energy. Although previously dismissed, an LBH is a potential candidate in accounting for the 1908 devastation of Tungus Siberia, since important LBH interactions were overlooked. LBH passing through neutron star pulsars are capable of causing a sudden change in frequency which may not be fully accounted for by other theories. Rapid energy input due to the passage of LBH through the earth, sun, and neutron stars is examined to determine if they could initiate tremors and quakes in such bodies. A case is made that in encounters with the earth's atmosphere, cosmic LBH can manifest themselves as the core energy source of ball lightning (BL). Relating the LBH incidence rate on earth to BL occurrence has the potential of shedding light on the distribution of LBH in the universe, and their velocities relative to the earth. Most BL features can be explained by a testable LBH model. Analyses are presented to support this model. The total number of degrees of freedom of a d-dimensional body in n-space is derived so that equipartition of energy may be applied in the early universe and related to LBH. Blackbody and Hawking radiation are generalized to n-space. The entropy of LBH and of the universe are examined. Novel black hole entropy equations are obtained, which may shed light on the enigma of why the primordial universe appears to have so extremely little entropy. The largest possible attractive force, repulsive force, and luminosity in nature are considered in the context of LBH. A question is raised as to whether the Planck scale is truly fundamental. The gravitational fine structure constant is re-examined.

**Chapter Sections**

1  Introduction
2  Radiation in n-Space

# 1 Introduction

The discovery of the accelerated expansion of the universe was totally unanticipated. It calls long-standing cosmological theories into question. It introduces new enigmas related to dark matter/dark energy; and may shed light on an old enigma that some stars appear to be older than the previously accepted age of the universe. The implication is that the universe is older, bigger, and less dense than previously thought by a number of accepted measurements. The new discoveries radically change our concept of what drives the macrocosm, and initiate a fundamentally new quest for the laws that govern the universe on a large scale.

Our universe is so full of surprises that caution should always be the byword. An old saying has it that cosmologists proceed undaunted in being almost always certain, but rarely right. This has been the case with regards to the character of the expansion of the universe. It has long been taken for granted that the expansion of the universe is either at a constant rate, or decelerating due to the gravitational attraction of all the mass in it. So it came as quite a surprise in 1998 when two independent international groups of astrophysicists at Lawrence Berkeley National Lab (Perlmutter *et al.*,1998) in the U.S., and Mount Strombo and Siding Spring Observatories (Riess *et al.*, 1998) in Australia, using type Ia supernovae to gauge distances, discovered that the universe is accelerating in its expansion. One viable competing explanation is that accelerated expansion of the universe is due to radiation from little black holes (LBH) propelling them outward and gravitationally towing ordinary matter with them. Little black holes may be the dark matter/dark energy representing 95% of the mass of the universe (Rabinowitz, 1998, 1999 a,b, 2001a,b,c, 2003).

## 2 Radiation in n-Space

Understanding the distinction between Hawking radiation and Gravitational Tunneling Radiation from black holes is key to being able to comprehend and discriminate between various views of black holes as dark matter candidates. Since Hawking radiation is presented as blackbody radiation from a black hole, ordinary blackbody radiation is a good starting point. Because so much of Modern Cosmology involves the possibility of higher dimensional space, let us develop equations in n-space which can easily be reduced to their 3-space counterparts.

### 2.1 Blackbody radiation in n-space

Let us generalize Boltzmann's derivation of the blackbody radiation law. In n-space, the radiation pressure $P_n = \frac{1}{n} u_n$, where $u_n$ is the energy density of the radiation. The internal energy $U_n = u_n V_n$, where $V_n$ is the n-volume. The thermodynamic relation for internal energy is

$$\frac{\partial}{\partial V_n}(U_n)_T = T\left(\frac{\partial P_n}{\partial T}\right)_V - P_n \Rightarrow \frac{\partial}{\partial V_n}(u_n V_n) = T\frac{\partial}{\partial T}\left(\frac{u_n}{n}\right) - \frac{u_n}{n}. \tag{2.1}$$

Equation (2.1) leads to

$$\frac{du_n}{u_n} = (n+1)\frac{dT}{T} \Rightarrow u_n \propto T^{n+1}. \tag{2.2}$$

Thus the n-dimensional equivalent of the Stefan-Boltzmann blackbody radiation law from eq. (2.2) is

$$P_{BBn} \propto c u_n \propto T^{n+1}, \tag{2.3}$$

***It is interesting to note that the dimensionality of macroscopic space can be determined by measuring the exponent of the blackbody radiation law.*** If energetically stable atoms (e.g. bound by additional short-range forces) could exist in (n > 3)-space (cf. Sec. 18), eq. (2.3) says that for high T, the collective blackbody radiation of these atoms emits considerably higher power than in 3-space, and for low T the opposite is true.

### 2.2 Hawking radiation in n-space

The Hawking radiated power, $P_{SH}$, follows from the Stefan-Boltzmann blackbody radiation power/area law $\sigma T^4$ for black holes. For Hawking (1974, 1975):

$$P_{SH} \approx 4\pi R_H^2 \left[\sigma T^4\right] = 4\pi \left(\frac{2GM}{c^2}\right)^2 \sigma \left[\frac{\hbar c^3}{4\pi kGM}\right]^4 = \frac{\hbar^4 c^8}{16\pi^3 k^4 G^2} \{\sigma\} \left[\frac{1}{M^2}\right] \quad (2.4)$$

where $\sigma$ is the Stefan-Boltzmann constant. $R_H = 2GM/c^2$ is the Schwarzchild radius, often also called the horizon of the black hole. To avoid the realm of quantum gravity, Hawking requires the black hole mass $M > M_{Planck}$.

Since Hawking radiation was developed as blackbody radiation from a black hole, using eq. (2.3), $R_{Hn}$ and $T_n$ from (Rabinowitz, 2001a, b), the Hawking power radiated in n-space for $n \geq 3$:

$$P_{SHn} \propto \left[R_{Hn}\right]^{n-1} \left[T_n\right]^{n+1} \propto \left[M^{1/(n-2)}\right]^{n-1} \left[M^{-1/(n-2)}\right]^{n+1} \propto \frac{1}{M^{2/(n-2)}}$$

$$\propto M^{-2} \text{ for 3 – space.} \quad (2.5)$$

$$\propto M^{-1/4} \text{ for 10 – space.}$$

Although ordinary blackbody radiation is dramatically large in 4-space and higher e.g. $\propto T^{11}$ in 10-space, the mass dependency of Hawking radiation causes it to decrease for dimensions higher than 3 for LBH.

**2.3 Compact dimensions attenuate Hawking radiation**

Another approach assumes the correctness of the Hawking model, but analyzes the effects of additional compact dimensions on the attenuation of this radiation. Argyres et al (1998) conclude that the properties of LBH are greatly altered and LBH radiation is considerably attenuated from that of Hawking's prediction. Their LBH are trapped by branes so essentially only gravitons can get through the brane (which may be thought of as an abbreviation for vibrating membrane). For them, not only is the radiation rate as much as a factor of $10^{38}$ lower than given by Hawking, but it also differs by being almost entirely gravitons.

**2.4 Gravitational tunneling radiation (GTR)**

Gravitational tunneling radiation (GTR) may be emitted from black holes in a process differing from that of Hawking radiation, $P_{SH}$, which has been undetected for over three decades. Belinski (1995), a noted authority in the field of general relativity, unequivocally concludes "the effect [Hawking radiation] does not exist." GTR is offered as an alternative to $P_{SH}$. In the gravitational tunneling model (Rabinowitz,1999 a,b), beamed exhaust radiation tunnels out from a LBH with radiated power, $P_R$, due to the field of a second body, which lowers the LBH gravitational potential energy barrier and gives the barrier a finite width. Particles can escape by tunneling (as in field emission).

This is similar to electric field emission of electrons from a metal by the application of an external field.

Although $P_R$ is of a different physical origin than Hawking radiation, we shall see that it is analytically of the same form, since $P_R \propto \Gamma P_{SH}$, where $\Gamma$ is the transmission probability approximately equal to the WKBJ (Wentzel-Kramers-Brilloin-Jefferies, also called WKB) tunneling probability $e^{-2\Delta\gamma}$ for LBH. The tunneling power radiated from a LBH for $r \gg R_H$ is:

$$P_R \approx \left[\frac{\hbar c^6 \langle e^{-2\Delta\gamma}\rangle}{16\pi G^2}\right]\frac{1}{M^2} \sim \frac{\langle e^{-2\Delta\gamma}\rangle}{M^2}\left[3.42\times 10^{35} W\right], \qquad (2.6)$$

where M in kg is the mass of the LBH. No correction for gravitational red shift needs to be made since the particles tunnel through the barrier without change in energy. The tunneling probability $e^{-2\Delta\gamma}$ is usually $\ll 1$ and depends on parameters such as the width of the barrier, M, and the mass of the second body (Rabinowitz,1999 a,b).

Hawking invoked blackbody radiation in the derivation of eq. (2.4). But it was not invoked in the GTR derivation of eq. (2.6). Although $P_R$ and $P_{SH}$ represent different physical processes and appear quite disparate, the differences in the equations almost disappear if we substitute into eq. (2.4) the value obtained for the Stefan-Boltzmann constant $\sigma$ by integrating the Planck distribution over all frequencies:

$$\sigma = \left\{\frac{\pi^2 k^4}{60\hbar^3 c^2}\right\}, \qquad (2.7)$$

$$P_{SH} = \frac{\hbar^4 c^8}{16\pi^3 k^4 G^2}\left\{\frac{\pi^2 k^4}{60\hbar^3 c^2}\right\}\left[\frac{1}{M^2}\right] = \frac{\hbar c^6}{16\pi G^2}\left\{\frac{1}{60}\right\}\left[\frac{1}{M^2}\right]. \qquad (2.8)$$

Thus $P_R = 60\langle e^{-2\Delta\gamma}\rangle P_{SH}$. $\qquad (2.9)$

GTR is beamed between a black hole and a second body, and is attenuated by the tunneling probability $\langle e^{-2\Delta\gamma}\rangle$ compared to $P_{SH}$. Leaving aside the attenuation factor, $\langle e^{-2\Delta\gamma}\rangle$, it is not clear if there is physical significance to the same analytic form for $P_R$ and $P_{SH}$. It could simply result from the dimensionality requirement that they are both in units of power.

Two LBH may get quite close for maximum GTR. In this limit, $\langle e^{-2\Delta\gamma}\rangle \to 1$, and there is a similarity between GTR and what is expected from Hawking's model. GTR produces a repulsive recoil force between two bodies due to the beamed emission between them. Since the tidal forces of two LBH add together to give more radiation at their interface in his model, this also produces a repulsive force.

**2.5 Diminished Hawking radiation from charged black holes**

As a black hole becomes more and more charged, the Hawking radiation decreases until in the limit of maximum charge containment there is none. Balbinot (1986) demonstrated that highly charged black holes do not Hawking radiate. He determined that "For an extreme Reissner-Nordstrom [highly charged] black hole ... there is no Hawking evaporation." There is no mention of this in Chavda and Chavda (2002) whose atomic black hole model will be discussed in Sec. 3. They did not address the question of the accelerated expansion of the universe. Nor do they consider moderately charged black holes which could form electrostatically and gravitationally bound atoms. Chavda and Chavda (2002) realized that the universe can be better understood without Hawking radiation. However it appears that they used an incorrect method to dispense with it. Though charged black holes have not been candidates for dark matter, charged black hole atoms, neutralized by orbiting charges, have been considered by Rabinowitz (1999 a).

**2.6  Gravitational radiation**

An advanced quadrupole suspension design (Robertson, 2002) for the U.S. Laser Interferometer Gravitational-Wave Observatory (LIGO) has recently been described to measure gravitational radiation from distant sources. Preparation is also being made by other teams around the world. In addition to LIGO, there is VIRGO (France/Italy); GEO-600 (Britain/ Germany); TAMA (Japan); and ACIAGA (Australia). The detectors are laser interferometers with a beam splitter and mirrors suspended on wires. The predicted gravitational wave displaces the mirrors and shifts the relative optical phase in two perpendicular paths. This causes a shift in the interference pattern at the beam splitter. It is expected that by 2010, the devices will be sensitive enough to detect gravitational waves up to $10^2$ Megaparsecs ( $3.26 \times 10^8$ lightyear = $3.1 \times 10^{24}$ m) away. A perplexity arose because the detector noise does not satisfy the usual assumptions that it be stationary and Gaussian (Allen et al, 1999)

This perplexity may be due to radiation from gravitationally bound atoms (GBA). As shown in (Rabinowitz, 2001a,b, 2003), quantized gravitational radiation is possible from GBA. The possibility was presented that a signal from such potentially nearby sources can compete or interfere with distant sources such as neutron stars, binary pulsars, and coalescing black holes. Signals from such distant sources are expected to have frequencies in the range from 10 Hz to $10^4$ Hz (Davies, 1992). It was shown that gravitational radiation from orbital de-excitation of an ordinary mass orbiting a LBH would have a detectable frequency ~ $10^3$ Hz [2]. A mass m ~ $10^{-27}$ kg orbiting a LBH of mass M ~ 10 kg, would emit a frequency ~ $10^3$ Hz. in going from the j = 3 state to the j = 1 ground state.

# 3  Different Views of Black Holes as Dark Matter Candidates

Discovery of the nature of dark matter/dark energy will help to define what the universe is made of. It will reveal the invisible particles carrying the gravitational glue that holds the universe, galaxies, and clusters of galaxies together, and determines the curvature of space. We should not arbitrarily rule out the possibility that dark matter can occasionally display itself on earth; and we shall explore what form its manifestation may take. Let us briefly look at various views of black holes as dark matter candidates to give us a broad perspective.

### 3.1 Large black holes: $10^{14}$ kg ≤ $M_{BH}$ ≤ $10^{36}$ kg

A 1984 review article (Blumenthal et al and references therein) presents the then and presently prevailing view of black holes as constituents of dark matter. The article considers only rather massive black holes as a possible component of dark matter: "A third cold DM [dark matter] candidate is black holes of mass $10^{-16}$ $M_{sun}$ ≤ $M_{BH}$ ≤ $10^6$ $M_{sun}$, the lower limit implied by the non-observation of γ rays from black hole decay by Hawking radiation...." $M_{sun}$ = 2 x $10^{30}$ kg implies that for them $10^{14}$ kg ≤ $M_{BH}$ ≤ $10^{36}$ kg.

### 3.2 Medium black holes: $10^{12}$ kg ≤ $M_{BH}$ ≤ $10^{30}$ kg

Trofimenko (1990) discussed the possibility that black holes up to the mass of the sun, $M_{sun}$, are involved in a multitude of geophysical and astrophysical phenomena such as in stars, pulsars, and planets. Although he did not explicitly consider them as candidates for dark matter, for him they are "universal centres of all cosmic objects." That implicitly makes them dark matter candidates . He was not concerned with the ramifications of LBH radiation, nor the time for LBH to devour their hosts. His lower mass limit of $10^{12}$ kg comes from the failure to detect Hawking radiation, and expected smallest primordial mass survival.

### 3.3 Primordial black holes: $M_{BH}$ ~ $10^{13}$ kg

Alfonso-Faus (1993) proposed "primordial black holes, massive particles about $10^{40}$ times the proton mass" [$10^{40}(10^{-27}$ kg) = $10^{13}$ kg] as his dark matter candidate. He goes on to say that they do not radiate by Hawking radiation, but did not then comment on how they radiate. Later Alfonso-Faus (1999) asserts a radiation wavelength of $10^8$ cm from black holes that is the geometric mean between the radius of such a primordial black hole ($10^{-12}$ cm) and the radius of the universe ($10^{28}$ cm). With such a long wavelength, he concludes that they radiate, "about $10^{40}$ times lower " than in the Hawking model and hence "they would still be around....."

### 3.4. Primordial black holes in higher dimensional space: $10^{29}$ kg ≤ $M_{BH}$ ≤ $10^{34}$ kg

Argyres et al (1998) examine primordial black holes (PBH) in higher compact dimensions. They conclude that for 6 extra compact dimensions (9-space), 0.1 solar mass PBH are dark matter candidates, but that this increases to ~$10^4$ solar masses if there are only 2 or 3 extra dimensions (5 to 6-space). So for them $10^{29}$ kg ≤ $M_{BH}$ ≤ $10^{34}$ kg. Smaller PBH might be expected to abound, since for them PBH radiation is almost entirely gravitons. In standard Hawking radiation from LBH, > MeV photons would dissociate big bang nucleosynthesis products, devastating the presently propitious predictions of light element abundances. They conclude, "The lightest black holes that can be present with any significant number density in our universe today are thus formed immediately after the epoch of inflationary reheating."

## 3.5  GTR radiating primordial little black holes: $10^{-7}$ kg ≤ $M_{LBH}$ ≤ $10^{19}$ kg

Starting in 1998, this author proposed that black holes radiate by gravitational tunneling radiation (GTR) resulting in a most compelling case for primordial LBH as the main constituent of dark matter of the universe (Rabinowitz, 1998 a,b, 1999, 2001a,b, 2003). These were the smallest masses ($10^{-7}$ kg to $10^{19}$ kg) considered until 2002. Since GTR is greatly attenuated compared with Hawking radiation, cf. Section 6.4, this has strong implications down to the smallest masses of LBH, whether the LBH are free or are gravitationally bound atoms. For Hawking (1974, 1975), the smallest LBH that can survive to the present is $M \sim 10^{12}$ kg.

The GTR model is only briefly covered in this review section, since its implications are further examined elsewhere in this Chapter. Let us look here at one of the predictions of GTR. The evaporation rate for a black hole of mass M is $d(Mc^2)/dt = -P_R$, which gives the lifetime

$$t = \frac{16\pi G^2}{3\hbar c^4 \langle e^{-2\Delta\gamma} \rangle} [M^3]. \tag{3.1}$$

This implies that the smallest mass that can survive up to a time t is

$$M_{small} = \left( \frac{3\hbar c^4 \langle e^{-2\Delta\gamma} \rangle}{16\pi G^2} \right)^{1/3} [t^{1/3}]. \tag{3.2}$$

Primordial black holes with $M \gg M_{small}$ have not lost an appreciable fraction of their mass up to the present. Those with $M \ll M_{small}$ would have evaporated away long ago.

Thus the smallest mass that can survive within ~ $10^{17}$ sec (13.7 x $10^9$ year = age of our universe) is

$$M_{small} \geq 10^{12} \langle e^{-2\Delta\gamma} \rangle^{1/3} \text{ kg}. \tag{3.3}$$

Hawking's result (1974, 1975) of $10^{12}$ kg is obtained by setting $e^{-2\Delta\gamma} = 1$ in eq. (3.3). Since $0 \leq e^{-2\Delta\gamma} \leq 1$, an entire range of black hole masses much smaller than $10^{12}$ kg may have survived from the beginning of the universe to the present than permitted by Hawking's theory.

For example, if the average tunneling probability $\langle e^{-2\Delta\gamma} \rangle \sim 10^{-45}$, then $M_{small} \sim 10^{-3}$ kg. For $M_{univ} \sim 10^{53}$ kg, $V_{univ} \sim 10^{79}$ m$^3$ (radius of 14 x$10^9$ light-year $\approx 1.4 \times 10^{26}$ m), the average density of such LBH would be 1 LBH per $10^{23}$m$^3$. The velocity of our local group of galaxies with respect to the microwave background (cosmic rest frame), 6.2 x $10^5$ m/sec (Turner and Tyson, 1999), is a reasonable velocity for LBH with respect to the earth. This may make it possible to detect their incident flux $\sim (10^{-23}/m^3)(6.2 \times 10^5 \text{ m/sec}) \sim 10^{-17}/m^2\text{sec}$ on the earth (Rabinowitz, 2001 a,b), about which we will go into more detail in this Chapter.

### 3.6 Non-radiating holeum : $10^{-24}$ kg $\leq M_{BH} \leq 10^{-12}$ kg

By analogy with the neutron, Chavda and Chavda (2002) introduced a novel proposal that gravitationally bound black holes will not Hawking radiate. Free neutrons are unstable, but bound neutrons are stable in most nuclei. Their model is briefly reviewed in this section, and will be further analyzed in detail in Secs, 16 and 17, as it dispenses with Hawking radiation in a novel way. It appears from my analysis that stable holeum cannot exist in 3-space, or in any higher dimensions. Therefore whether or not such an object might Hawking radiate is a moot point. They consider the range $10^{-24}$ kg $\leq M_{BH} \leq 10^{-12}$ kg.

The analogy between holeum and a bound neutron may not apply. A neutron in free space decays with a half-life of about 10.6 minutes. The neutron spontaneously decays into a proton, an electron, and an antineutrino. This is energetically possible because the neutron's rest mass is greater than that of the decay products. This difference in rest mass manifests itself in an energy release of 1.25 x $10^{-13}$ J (0.782 MeV). The situation in a nucleus is complicated by many factors such as Fermi levels of the neutrons and the protons, etc. Neutrons do decay in nuclei that are beta emitters despite their relatively large binding energy which is typically 1 to 1.4 x $10^{-12}$ J (6 to 8 MeV). Other than the interesting neutron analogy, they give no compelling reasons for the absence of Hawking radiation in black hole GBA.

Most of the orbital radii are in the strong field region 2 $R_H$ < r < 10 $R_H$, requiring general relativity corrections. Therefore in the absence of r > 10$R_H$, their use of Newtonian gravity is questionable. There is an error by a factor of $10^2$ too high in the orbital radius given by their eq. (45).

In considering little black hole masses $M_{LBH}$ < $M_{Planck} \sim 10^{-8}$ kg, their analysis exceeds another domain of validity which requires a theory of quantum gravity. For larger masses and larger radii than they use, it would be easy to agree with their choice of quantized Newtonian Gravity.  Except for the 0 angular momentum state (which does not exist semi-classically), essentially the same results are obtained semi-classically as are gotten quantum mechanically for hydrogen.  It is generally agreed that for $M_{LBH}$ > $M_{Planck}$ one may describe LBH with semi-classical physics, and quantum gravity is needed for $M_{BH} \leq 10^{-8}$ kg, since this is below the Planck scale where a little black hole has $R_H \leq 10^{-35}$ m.

Neglect of special relativity is a further problem, since in some cases the orbital velocity in holeum $v \approx c$.  It is relevant to note that non-relativistic quantum mechanics and even the semi-classical Bohr-Sommerfeld equation give accurate energy levels for hydrogen despite being non-relativistic.  This is because the serendipitously near-canceling effects of both relativity and spin are neglected.  One effect is the relativistic increase of the electron's mass as its velocity increases near the proton.  The other effect is the interaction of the electron's intrinsic magnetic moment with the Coulomb field of the proton.  Since a neutral LBH has no magnetic moment, there are no canceling effects and one may expect a much less reliable result from a treatment which ignores special relativity, such as theirs.

## 4  Degrees of Freedom and Equipartition of Kinetic Energy in n-Space

The equipartition of kinetic energy is an important principle that enters into many analyses related to the Universe, as well as those in this Chapter.  With the possibility of higher dimensional space it would be useful to have a generalization of it to any dimension n.  Since I could not find it in the literature, here is my generalization

### 4.1  Degrees of freedom in n-space

The total number of degrees of freedom $D_n$ of a d-dimensional body in n-space is

$$D_n = n + (n-1) + (n-2) + ... + (n-d), \tag{4.1}$$

for $d \leq n$.  Once n coordinates establish the center of mass, there are (n - 1) coordinates left to determine a second reference point on the body, leaving (n - 2) for the third point, ..., and finally (n - d) coordinates for the (d + 1)th reference point.

Since the RHS of eq. (4.1) has (d + 1) terms:

$$\begin{aligned} D_n &= n + (n-1) + (n-2) + ... + (n-d) = (d+1)\left(\frac{n+(n-d)}{2}\right) \\ &= \left(\frac{d+1}{2}\right)(2n-d). \end{aligned} \tag{4.2}$$

In 3-space, for example, both a 2-dimensional object and a 3-dimensional object each have 6 degrees of freedom.

It is interesting to note that the latter observation is true in general i.e. $D_n$ is the same for $d = (n - 1)$ and for $d = n$:

$$D_n(d = n, \text{or } n-1) = \left(\frac{n+1}{2}\right)(2n - n) = \left(\frac{(n-1)+1}{2}\right)[2n - (n-1)] = \frac{n(n+1)}{2} \quad (4.3)$$

**4.2 Generalization of equipartition of kinetic energy to n-space**

In 3-space, $D_3$ varies from 3 for $d = 0$ (point-like object) to 6 for $d = 2$ (planar object like an ellipse) or $d = 3$ (object like a spheroid). Choosing $n = 10$ in reference to 10-space string theory, eq. (4.2) shows that $D_{10}$ varies from 10 for $d = 0$, to 55 for $d = 9$ or 10 i.e. a 9 or 10-dimensional object.

Because the kinetic energy is a quadratic function of velocity in n-space, there will be on the average $(1/2)kT$ of kinetic energy per degree of freedom $D_n$. Let us consider two cases:

1) 3-dimensional body (which could be bound by short range forces) in n-space, i.e. $d = 3$;

2) n-dimensional body in n-space, i.e. $d = n$, where n is the number of spatial dimensions in the space-time manifold of $(n+1)$ dimensions.

For a 3-dimensional body in n-space, from eq. (4.2) the average kinetic energy is

$$\langle KE \rangle = D_n\left(\tfrac{1}{2}kT\right) = \left(\frac{3+1}{2}\right)(2n-3)\left(\tfrac{1}{2}kT\right) = (2n-3)kT \quad (4.4)$$

$= 17\, kT$ for $n = 10$. [5 kT for a point-like body, depending on scale.]

For an n-dimensional body in n-space, using eq. (4.4) gives

$$\langle KE \rangle = D_n\left(\tfrac{1}{2}kT\right) = \left(\frac{n+1}{2}\right)(2n-n)\left(\tfrac{1}{2}kT\right) = \left(\frac{n(n+1)}{4}\right)kT. \quad (4.5)$$

$= 3\, kT$ for $n = 3$. [(3/2)kT for a point-like body, depending on scale.]

For $n = 10$, a 10-dimensional body has $\langle KE \rangle = (55/2)kT \approx 28\, kT$.

Thus in terms of equipartition of energy, at a given temperature T, there can be significantly higher kinetic energy than expected in higher dimensions. Chavda and Chavda (2002) are interested in the early universe when the temperature $T \gg T_b \equiv mc^2/k$, where m is each black hole mass which makes up holeum. In 3-space, the average kinetic energy is between 3/2 and 3 $mc^2$ depending on the scale of interaction as to whether the black holes should be considered point-like or 3-dimensional in collisions, and a large percentage of holium collisions may result in dissociation. If extra dimensions are unfurled in the early universe, from eqs. (4.4) and (4.5), in 10-space, holeum (if also bound by short range forces) would clearly be dissociated in collisions, since the average kinetic energy would be as high as 5 to 28 mc$^2$. Additional problems with their model will be presented later in this Chapter.

## 5 Little Black Holes and Ball Lightning

Prior to the awareness that LBH radiate appreciably, their presence on earth was considered highly unlikely, as it was expected that LBH would devour the earth in times ~ million years. But with Hawking radiation evaporation of LBH, their lifetime < ~ year would be much less than the time it would take to ingest the earth. However, LBH would be unlikely on earth with Hawking radiation, because this devastating radiation in all directions has not been observed. The view of radiation from LBH presented by the Rabinowitz model (1999 a, b) obviates both of the above problems since this radiation is beamed and considerably less than Hawking's (1974, 1975). In the Rabinowitz model, when LBH get so small that there would be appreciable rocket-like exhaust radiation, the radially outward radiation reaction force propels them away from the earth like a rocket ship.

Ball lightning is widely accepted, but still unexplained. A testable LBH model for BL is presented which explains most of the known features of BL. In this model, LBH produce visible light in interacting with the atmosphere. The BL core energy source is gravitationally stored energy which is emitted as beamed radiation by means of gravitational field emission.

Most of the results in Secs. 6 - 9 are derived independently of the model of black hole radiation. Near the LBH, exhaust radiation can augment ionization and excitation, but this complication will not be introduced at this time. Although a number of mechanisms are at work, orbital trapping with subsequent polarization and ionization by the LBH gravitational and electrostatic tidal force is the major direct LBH ionization mechanism. LBH with mass < ~ 10$^{-3}$ kg and radius <~ 10$^{-30}$ m are found to be the most likely candidates to manifest themselves as ball lightning (BL).

## 6 LBH Gravitational and Electrostatic Tidal Force

### 6.1 Gravitationally enhanced ionization cross-section

The intense attractive converging gravitational and/or electrostatic field of a charged LBH causes more atmospheric molecules to be polarized and ionized than given by only kinetic considerations. Let us first examine the gravitational case. The gravitational potential energy of a particle of mass m in the field of a LBH is

$$V = -\frac{GMm}{r} - p\left(\frac{GM}{r^2}\right) - \frac{\alpha_p}{2}\left(\frac{GM}{r^2}\right)^2, \qquad (6.1)$$

where p is the permanent dipole moment, which will usually be negligible for atoms but not for molecules, and $\alpha_p$ is the gravitational polarizability. We will be dealing primarily with atoms of the disassociated molecule since the binding energy of the molecules << the ionization potential, and they will be torn apart well before getting in close enough for ionization. When the atomic collision frequency is low compared with the ionization rate due to tidal interaction with the gravitational field of the LBH, the ionization radius $r_i$ can be increased. This results in an enhanced ionization volume, i.e. an enhanced ionization cross-section $\sigma_E$.

To a first approximation, this problem will be treated as a simple central force problem in which angular momentum is conserved. Implications of (1) atomic scattering, (2) ionization and scattering by the LBH exhaust, and (3) tidal force interactions will be neglected for now. These make orbital motion non-reentrant about the LBH as indicated by the Runge vector (or Runge-Lenz vector, quantum mechanically). Scattering is negligible as an LBH enters the low density atmosphere from outer space and starts to produce ions around it, and as we shall see even at high density, when the mean free path $\lambda$ > the enhanced interaction radius as calculated in this section. The interaction analysis here applies to both the sphere of ionization and to the sphere of polarization. So the symbol $r_{ip}$ will represent either the ionization radius or the larger polarization radius depending on which case is to be considered.

We can make the problem one-dimensional involving only the radial dimension, by introducing an effective potential energy

$$V_{eff} = V(r) + \frac{L^2}{2mr^2}, \tag{6.2}$$

where L is the conserved angular momentum of an atom about the LBH.

$$L = mvr_E = m(\Im kT/m)^{1/2} r_E = (\Im mkT)^{1/2} r_E, \tag{6.3}$$

where T is the temperature of the gas, $\Im \approx 3$ is the number of degrees of freedom of the particle, and $r_E$ is the enhanced ionization radius, i.e. the larger orbit capture radius of furthest approach for ionization of an atom.

The radial velocity $v_r = [2(E - V_{eff})/m]^{1/2} = 0$ at the closest approach to $r_i$, for a particle that just grazes the original ionization sphere. Hence at $r = r_{ip}$, $V_{eff} = E = \frac{\Im}{2} kT$. Combining this with eqs. (6.1) and (6.2) yields

$$V(r_{ip}) = \tfrac{\Im}{2} kT \left[ 1 - (r_E/r_{ip})^2 \right]. \tag{6.4}$$

Therefore eq. (6.4) gives us the enhanced ionization radius,

$$r_E = r_{ip} \left[ 1 - \frac{1}{\tfrac{\Im}{2} kT} V(r_i) \right]^{1/2}. \tag{6.5}$$

The ionization-polarization radius increases since V is negative, i.e. it is attractive.

We next need to determine the gravitational polarizability $\alpha_p$. The gravitational tidal force $F_T$ polarizes an atom,

$$F_T = \frac{(ze)^2 \partial}{4\pi \varepsilon a^3}, \tag{6.6}$$

-13-

where a is the unperturbed atomic radius, $\partial$ is the displacement relative to the electron cloud of the nucleus of mass $m_N \approx m$ the atomic mass, and $\varepsilon$ is the permittivity of free space. The term on the right is the electrical harmonic restoring force with spring constant $K = (ze)^2/4\pi\varepsilon a^3$. The displacement produces both electric and gravitational dipole moments, of which the latter is

$$m_N \partial \approx m \partial = \alpha_p (F_T/m). \tag{6.7}$$

Combining eqs. (6.6) and (6.7) yields a general result independent of the form of $F_T$.

$$\alpha_p = \left(\frac{4\pi\varepsilon m^2}{(ze)^2}\right) a^3. \tag{6.8}$$

Substituting eq. (6.8) into (6.1),

$$\begin{aligned} V &= -\left(\frac{GMm}{r}\right) - p\left(\frac{GM}{r^2}\right) - \frac{1}{2}\left(\frac{4\pi\varepsilon m^2 a^3}{(ze)^2}\right)\left(\frac{GM}{r^2}\right)^2 \\ &= -\left(\frac{GMm}{r}\right) - p\left(\frac{GM}{r^2}\right) - \left[2\pi\varepsilon\left(\frac{GMm}{ze}\right)^2\right]\frac{a^3}{r^4} \end{aligned} \tag{6.9}$$

Thus from eq. (6.5), the gravitational orbit capture enhanced ionization cross-section of the LBH is

$$\sigma_{gE} = \pi r_{ip}^2 \left\{1 + \frac{1}{\frac{3}{2}kT}\left[\frac{GMm}{r_{ip}} + p\left(\frac{GM}{r_{ip}^2}\right) + 2\pi\varepsilon\left(\frac{GMm}{ze}\right)^2 \frac{a^3}{r_{ip}^4}\right]\right\}. \tag{6.10}$$

Equation (6.10) is applicable if the particle mean free path $\lambda < r_E$. For convenience, this will be called the low density case. Whether the low or high density case is relevant is a function of both the density of the gas and the mass M of the LBH, since for small $r_E$, the mean free path $> r_E$ even above atmospheric pressure.

**6.2 Electrostatically enhanced ionization cross-section**

A similar analysis with analogous steps can be done for the electrostatic case which can dominate over the gravitational case. The resulting electrostatic orbit capture enhanced ionization cross-section of a charged LBH is

$$\sigma_{eE} = \pi r_{ip}^2 \left\{1 + \frac{1}{\frac{3}{2}kT}\left[\frac{Qq}{4\pi\varepsilon r_{ip}} + p\left(\frac{Q}{4\pi\varepsilon r_{ip}^2}\right) + 2\pi\varepsilon\left(\frac{Q}{4\pi\varepsilon}\right)^2 \frac{a^3}{r_{ip}^4}\right]\right\}, \tag{6.11}$$

where Q is the electric charge of the LBH, q is the net charge of the atom or molecule, and the electric polarizability $\alpha_{pe} = 4\pi\varepsilon a^3$. Though only single ionization will be considered, higher degrees of ionization are possible.

Little black holes can also become visible indirectly as ball lightning in the surrounding air by excitation and direct collisional ionization with a charged little black hole resulting in electron ion pair recombinations, by excitation of the air molecules and atoms from the LBH exhaust radiation, and by infalling particle collisions.

## 7 Ball Lightning Radiation

### 7.1 Ionization rate

The ionization rate due to a LBH moving through the atmosphere is

$$\frac{dn_i}{dt} \sim \sigma \bar{v} n^2 + \sigma v_{BL} n^2 - \sigma_{re} \bar{v} n_i^2 - \sigma_{diff} \bar{v} n_i^2, \qquad (7.1)$$

where n is the number density of atoms, $\sigma$ is the ionization cross- section (enhanced or unenhanced depending on relative mean free path), $\sigma_{re}$ is the recombination cross-section, $\sigma_{diff}$ is the cross-section for diffusion out of the ionization sphere, $\bar{v}$ is the mean thermal velocity, $v_{BL}$ is the BL velocity, and $n_i$ is the number density of ions. The solution of eq. (7.1) is

$$n_i \sim \left[ \frac{A(Be^{Ct} - 1)}{(Be^{Ct} + 1)} \right]_{t \to \infty} \approx A = \left( \frac{\sigma(\bar{v} + v_{BL})n^2}{\sigma_t \bar{v}} \right)^{1/2} \qquad (7.2)$$

where $B = (A + n_{io})/(A - n_{io})$, $n_{io}$ is the initial number density of ions, $\sigma_t = \sigma_{re} + \sigma_{diff}$, and $C = 2n[\sigma \sigma_t (\bar{v} + v_{BL})\bar{v}]^{1/2}$.

### 7.2 Recombination radiation

As the LBH moves through the atmosphere, its gravitational and/or electrostatic tidal force excites and ionizes air atoms around it and carries the generated plasma along by electrostatic and/or gravitational attraction. At early times, the ionization time is short relative to the recombination time, and to the time for diffusion out of the ionization sphere. Entry into a LBH is difficult since the particle's deBroglie wavelength needs to be $< \sim R_H$, and because of conservation of angular momentum. In the presence of the LBH gravitational field and gradient, both the recombination and the diffusion times are longer than in free space, and $\sigma_{re}$ is reduced.

The equilibrium solution is obtained from eq. (7.2) as t gets large. In this limit the recombination rate is

$$R_{re} = [\sigma_{re} \bar{v} n_i^2] \sim \sigma(\bar{v} + v_{BL})n^2 \left\{ \left( \frac{\sigma \sigma_{re}}{(\sigma_{re} + \sigma_{diff})^2} \right) \left( \frac{\bar{v} + v_{BL}}{\bar{v}} \right) \right\}. \qquad (7.3)$$

The radiation is hardly perceptible at first. In steady state, the electron-ion recombination radiated power/volume is

$$P_{re} = [R_{re}]V_i \sim \left[\sigma(\bar{v}+v_{BL})n^2\left\{\left(\frac{\sigma\sigma_{re}}{(\sigma_{re}+\sigma_{diff})^2}\right)\left(\frac{\bar{v}+v_{BL}}{\bar{v}}\right)\right\}\right]V_i. \tag{7.4}$$

$V_i$ is the ionization potential (15.5 eV for nitrogen), and n is the number density of air atoms. For the LBH mass range of interest ~$10^{-3}$ kg, eq. (7.4) yields > ~ Watts of radiated power in agreement with observation. Photons with 15.5 eV energy have a frequency higher than visible photons of a few eV with wavelengths between 4000 Å and 8000 Å. However, energy degradation and other radiation mechanisms can result in visible light. There is comparable thermal and de-excitation radiation.

As will be derived in Sec. 12.1, a less detailed impulse transfer approach yields a power transfer of $P = \frac{4\pi G^2 M^2 \rho}{v_{bh}} \ell n\left(\frac{b_{max}}{b_{min}}\right) \sim 10$ W to the atmosphere by a LBH with M ~ $10^{12}$ kg, $R_H$ ~ $10^{-15}$ m, $\rho_{atm}$ = 1.3 kg/m$^3$ is the atmospheric mass density, and the weak logarithmic dependence of the ratio of the maximum to minimum impact parameters $\ell n(b_{max}/b_{min}) \sim 30$.

## 8 Beamed LBH Radiation Can Produce Levitation

The downwardly directed radiation (due to the earth below) from a LBH will act like a rocket exhaust permitting the LBH to levitate or fall slowly. Neglecting beam divergence, we can estimate the upward force on the LBH from

$$M\frac{dv}{dt} = -c\frac{dM}{dt} - Mg. \tag{8.1}$$

where the exhaust leaves the LBH at near the speed of light, c = 3 x $10^8$ m/sec, the acceleration of gravity g = 9.8 m/sec$^2$ near the earth's surface, and for levitation dv/dt = 0. The radiated power is related to the time rate of change of the LBH rest mass:

$$P_R = -\frac{dE}{dt} = -\frac{d}{dt}(Mc^2) = -c^2\frac{dM}{dt}. \tag{8.2}$$

Combining eqs. (8.1) and (8.2) gives the required radiated power for levitation in the earth's gravitational field as a LBH approaches the earth:

$$P_R = Mgc. \tag{8.4}$$

For a $10^{-7}$ kg LBH to levitate only $P_R$ ~ 300 W is needed for levitation; and for a 3 x$10^{-4}$ kg (1/3 gm) LBH to levitate, $P_R$ ~ $10^6$ W.

In one model, emission is mainly by the six kinds of neutrinos (Thorne et al., 1986) and in another almost entirely by gravitons (Argyres et al., 1998). The emitted power $P_R$, necessary to produce levitation, as well as the necessary masses and separations of the LBH and host body needed to produce this exhaust power are independent of the nature of the emitted particles. At a distance of many earth radii, the radiation is narrowly beamed toward the earth's center. As a LBH gets close to the earth the radiation beam diverges to ~ the earth's diameter, giving it a low power density.

## 9 Incidence Rate Of Ball Lightning

The continuity equation for mass flow of LBH when there is a creation rate $S_c$ and a decay rate $S_d$ of mass per unit volume per unit time t is

$$\nabla \bullet (\rho \vec{v}) + \partial \rho / \partial t = S_c - S_d, \tag{9.1}$$

where $\rho$ is the LBH mass density at a given point in the universe, $\vec{v}$ is the LBH velocity, and $\rho \vec{v}$ is the LBH flux density. In steady state, $\partial \rho / \partial t = 0$. Integrating eq. (9.1):

$$\int (\rho \vec{v}) \bullet d\vec{A} = \int (S_c - S_d) dV_t \Rightarrow$$
$$-\rho_{LBH} v_{LBH} A_{far} + \rho_{BL} v_{BL} A_E = (S_c - S_d) V_t \tag{9.2}$$

where $\rho_{LBH}$ is the mass density of LBH at a distance far from the earth, typical of the average mass density of LBH throughout the universe. $A_{far}$ is the cross-sectional area of a curvilinear flux tube of LBH far from the earth, $A_E$ is the cross-sectional area of the tube where it ends at the earth, and $V_t$ is the volume of the curvilinear flux tube (cylinder). Since the LBH were created during the big bang, at a large distance from the earth they should be in the cosmic rest frame. The velocity of our local group of galaxies with respect to the microwave background (cosmic rest frame), $v_{LBH}$ ~ 6.2 x $10^5$ m/sec (Turner and Tyson, 1999), is a reasonable velocity for LBH with respect to the earth. Interestingly, as shown in Sec. 11, this is also the escape velocity from our sun.

Because $v_{LBH}$ is high and LBH radiate little until they are near other masses, $S_c$ can be neglected with negligible decay of large black holes into LBH in the volume $V_t$. Similarly, $S_d$ may be expected to be small until LBH are in the vicinity of the earth where most of their evaporation, before they are repelled away, is in a volume of the atmosphere ~ $A_E h$, where $A_E$ is the cross-sectional area of the earth, and h is a characteristic height above the earth. At this point it is helpful to convert to number density $\rho_L$ and $\rho_B$, of LBH and ball lightning respectively. The number density decay rate is $\rho_B A_E h / \tau$, where $\tau < $ ~ year is the dwell-time of LBH near the earth. Thus eq. (9.2) yields

$$\rho_B = \rho_L \left[ \frac{v_{LBH}}{v_{BL} + (h/\tau)} \right] \frac{A_{far}}{A_E}, \tag{9.3}$$

which implies that the ball lightning flux is

$$\rho_B v_{BL} = \rho_L v_{LBH} \left[ \frac{v_{BL}}{v_{BL} + (h/\tau)} \right] \frac{A_{far}}{A_E} \approx \rho_L v_{LBH} \left( \frac{A_{far}}{A_E} \right), \tag{9.4}$$

where in most cases $h/\tau << v_{BL}$.

At large velocities, LBH that do not slow down appreciably due to their large mass or angle of approach, either do not produce sufficient ionization to be seen or do not spend sufficient time in the atmosphere to be observed. In the Rabinowitz model (1999a, b, c), those LBH that reach the earth's atmosphere and are small enough to have sufficient radiation reaction force to slow them down to the range of $10^{-2}$ to $10^2$ m/sec, with a typical value $v_{BL} \sim 1$ m/sec, manifest themselves as BL. So eq. (9.4) implies that the ball lightning current in the atmosphere $\approx$ the LBH current far away, i.e. $\rho_B v_{BL} A_E \approx \rho_L v_{LBH} A_{far}$. We can thus give a range for the BL flux density

$$\rho_L v_{LBH} < \rho_B v_{BL} < \rho_L v_{LBH} \left( \frac{A_{far}}{A_E} \right). \tag{9.5}$$

The distribution of LBH masses is not known. Assuming that LBH comprise all of the dark matter, i. e. 95 % of the mass of the universe (Rabinowitz, 1990 b) of which there is a percentage p of LBH of average mass $\overline{M}_{LBH} \sim 10^{-3}$ kg:

$$\rho_L \sim \frac{p(0.95 M_{univ} / \overline{M}_{LBH})}{V_{univ}}. \tag{9.6}$$

For $M_{univ} \sim 10^{53}$ kg, $V_{univ} \sim 10^{79}$ m$^3$ (radius of 15 x$10^9$ light-year = 1.4 x $10^{26}$ m), and p $\sim$ 10 %, $\rho_L \sim 10^{-24}$ LBH/m$^3$. [The critical density of the universe $\approx 10^{53}$ kg/$10^{79}$ m$^3$ = $10^{-26}$ kg/m$^3$ = $10^{-29}$ g/cm$^3$]. Thus from eqs. (9.5) and (9.6) my model predicts that the incidence rate of BL is roughly in the range

$$10^{-12} \text{ km}^{-2} \text{ sec}^{-1} \text{ to} >\sim 10^{-8} \text{ km}^{-2} \text{ sec}^{-1} \text{ for } A_{far}/A_E > \sim 10^4. \tag{9.7}$$

Even if p were 100%, $10^{-11}$ km$^{-2}$ sec$^{-1}$ would be well below the noise level of existing devices such as at large facilities for neutrino detection. This rate is in accord with the estimates for ball lightning incidence of Barry and Singer (1988) of 3 x $10^{-11}$ km$^{-2}$ sec$^{-1}$, and of Smirnov (1993) of 6.4 x $10^{-8}$ km$^{-2}$ sec$^{-1}$ to $10^{-6}$ km$^{-2}$ sec$^{-1}$. This is well below the incidence rate of ordinary lightning (Turman, 1977).

## 10  Meeting Ball Lightning /Earth Lights Criteria

If greatly decreased radiation permits little black holes to be prevalent throughout the universe, then it is reasonable to surmise that they are also present in the region of the earth. If they are present on earth, then one may ask how they might manifest themselves. If their presence can help to explain a long-known, well-established phenomenon that has no other explanation, then they are viable candidates for experimental investigations to test the validity of this hypothesis. It appears that ball lightning/earth lights represent an admirable testing ground.

A subtle variety of ball lightning are atmospheric luminous phenomena occurring in locations such as Hessdalen, Norway and elsewhere in the world. These are sometimes called "earth lights" (Devereaux, 1989), to make a refined distinction between them and ball lightning, as they appear to be more dynamic and unrelated to thunderstorm activity though otherwise they are very similar. At Hessdalen large numbers of researchers have observed earth lights moving parallel to the earth. This may just be a manifestation of little black holes where the LBH have approached somewhat horizontally or there is a large component of horizontal velocity due to a component of horizontal radiation reaction force because of the presence of mountains.

The Hessdalen sightings were visual, photographic, and had strong radar signals (Strand, 1984). Such observations are compatible with a charged levitating black hole. The luminosity and radar signals may be accounted for by the atmospheric ionization created by a charged little black hole and dragged along by electrostatic attraction to the hole. Lifetime measurements of the (ball lightning-like) earth lights at Hessdalen are among the most reliable as these were directly measured by numerous well-prepared observers both optically and with radar .

The following criteria are presented as a guide for assessing ball lightning/earth light models in general, and the little black hole model in particular. The first five are derived from Uman (1968), and the rest are inferred from different sources (Fryberger, 1994, and Singer, 1971).

**1)  Constant size, brightness, and shape for extended times**

The large amount of gravitationally stored energy in little black holes and resulting kinetic energy accounts for the somewhat constant size, shape, and brightness of ball lightning; and its particular shape is a function of the motion of the little black hole as it drags along ionized air. Ball lightning has stable spherical, pear-shaped, prolate and oblate ellipsoidal, cylindrical, and disk shapes (Singer, 1971).

Models that depend on thermally stored energy do not have stability due to cooling with time. For example, as given by eq. (3.1) a $<\sim 1/3$ gm little black hole can have a lifetime $\sim 1$ year near the earth. As it evaporates to a much smaller mass, with a concomitant increase in radiation reaction force, it will shoot up into space and thus extend its lifetime. Its luminosity can vanish when its trapped charge becomes neutralized, by going into the ground or other opaque structures, or when the black hole itself becomes disrupted, as possibly when the electrostatic repulsive force of the ingested charge ♠ the gravitational force that holds it together.

There are a number of models that fit this criterion. Finkelstein and Rubinstein (1964) proposed that ball lightning is a luminous region of air of nonlinear high electrical conductivity carrying a high current density. They showed that their model can yield ball-like solutions. A similar theory was presented by Uman and Helstrom (1966). Winterberg (1978) proposed an electrostatic theory of ball lightning.

**2)  Untethered high mobility**

The lightness of a little black hole ($<\sim 1/3$ gm) in which the ball lightning mass mainly resides, gives it high mobility. A small horizontal component of the exhaust force accounts for its horizontal mobility. A charged black hole will also experience an attractive force towards its image charge in a conductor, and either a repulsive or attractive force with a charged dielectric, depending on the relative sign of the charges.

Untethered mobility vitiates against electrical discharge models of ball lightning which require attachment to good (e.g., metal) or poor conductors (e.g., earth, wood) such as for St. Elmo's fire -- but attachment in either case.

**3) Generally doesn't rise**

The ball lightning ionized air is electrostatically bound to the charge trapped in the little black hole and so is forced to follow its trajectory rather than simply rise. Since heated air expands and rises, this is another criterion against thermal source ball lightning. Occasionally, ball lightning ascends faster than possible for heated air. Masses << 1/3 gm could rapidly ascend and vanish from the atmosphere. The majority of ball lightning observations are of a slow descent.

**4) Can enter open or closed structures**

The radius of a $3 \times 10^{-4}$ kg (1/3 gm) little black hole is $R_H = 2GM/c^2 = 5 \times 10^{-31}$ m. Uncharged little black holes have mean free paths through matter $>> 10^6$ km as shown in Sec. 13.2, and the mean free path of charged black holes $>>$ meters. Little black holes thus can easily penetrate through any material. Ohtsuki and Ofuruton (1991) have created plasma fireballs formed by microwave interference in air containing ethane and/or methane. These fireballs evidently can penetrate dielectric materials, but not metals. They may have difficulty meeting the requirement of low optical power. Smirnov (1990) and others have presented strong arguments that ball lightning cannot be a plasmoid. This criterion militates against most models that require external energy sources.

**5) Can exist within closed conducting metal structures**

Since little black holes have a more than adequate supply of stored energy they can easily exist inside any closed highly conducting structure. However, this criterion dictates against models that depend on electrical currents, microwaves, or other electromagnetic radiation that is shielded out by a conductor. Microwave models such as that of Kapitza (1968), Ohtsuki and Ofuruton (1991) and others would be ruled out in this case.

**6) Levitation**

The little black hole's downwardly directed radiation accounts for steady levitation. It is hard for other models to account for steady levitation while moving horizontally for long distances without rising.

**7) Low power in the visible spectrum**

In the example calculations of Sec. 8, although a little black hole may emit 300 W to $10^6$ W of total power outside the visible spectrum, it may only produce ~ 10 W of optical power by ionization of the surrounding air. The bulk of ball lightning observations (Singer, 1971) suggest that the observed intensities of light and heat are < ~ 10 W . This criterion rules out all those models for which the total visible radiated power would be far too great for the appropriate color temperature of the ball lightning.

**8) Rarity of sightings**

Almost everyone has seen lightning, but few people have seen ball lightning. Since little black holes are quite rare, this explains the rarity of sightings. Many models are not in accord with this criterion. With galactic concentration of the 95% black hole dark matter, ~ $10^3$ little black holes may be expected to be present in steady state in the region of the earth of volume $10^{12}$ km$^3$ (256 cubic billion miles) with ~ $1/10^9$ km$^3$ (~1 per cubic billion miles).

**9) Relatively larger activity near volcanoes**

Relatively larger activity of ball lightning near volcanoes has been reported. Given Trofimenko's proposal (1990) that LBH are the main source of heat for volcanoes, it follows that little black hole caused ball lightning should be more prevalent there. Other models don't explain this.

**10) Extinguishes quietly usually**

Ball lightning from little black holes usually extinguishes its luminosity quietly when it enters opaque materials like the ground or structures, slows down considerably, comes to rest, or becomes neutralized.

**11) Extinguishes explosively occasionally**

Ball lightning sometimes releases energy explosively (Singer, 1971). Little black holes occasionally extinguish explosively as their mass approaches $10^{-8}$ kg, or when otherwise disrupted. In 1846, lightning accompanied by fire-balls that "descended and exploded with terrific force" demolished the stone steeple of St. George's church in Leicester.

**12) Related radioactivity**

In examining the remains of the steeple apex mentioned in point 11), Mills (1971) looked for, but detected no radioactivity. He considered that radioactivity may have been undetectable because of the 125 years time lapse, but may have been detectable "within days of a ball lightning strike." Mills was testing the Altschuler et al (1970) model that ball lightning arises from a concentration of short-lived radioisotopes produced by lightning. There can be a low-level of $\gamma$-rays, positrons, and other radioactivity associated with ball lightning (Singer, 1993).

Ashby and Whitehead (1971) tested the hypothesis that ball lightning is caused by antimatter meteorites. They made radiation measurements over the period of one year near thunderstorms and tornadoes to check whether the annihilation of minute fragments of meteoric antimatter in the upper atmosphere could be the cause of ball lightning. Though radioactivity was detected, they seem to have disproved both the Altschuler et al hypothesis and their own model. Little black holes can account for radioactivity, whereas most other models cannot. See point 15) below regarding the recent finding of $\gamma$-radiation for long periods following lightning strikes.

**13) Typical absence of associated deleterious effects**

Because of the low interaction cross section of the emitted radiation and beam spreading, the total emitted GTR power from a little black hole has low local power deposition and low power density near the earth, dissipating over a large volume. As analyzed in Secs. 11 -13, most interactions in my GTR model of LBH radiation are not detrimental.

**14) Occasional high localized energy deposition**

Some ball lightning incidents require MJ of energy to account for molten materials and the reported boiling away of a bathtub full of water (Singer, 1971). The high energy content >> MJ of little black holes can account for this when a little black hole is disrupted by an end of life burst.

Dmitriev et al (1981) report an extremely high, well documented, localized energy deposition associated with an explosion observed by eye-witnesses including one of the authors: "The ball lightning was observed at 11:20 PM on August 23, 1978, in Khabarovsk, near Khazian Street, during a heavy rainfall. A suden whistle was heard similar to that produced by a jet engine. It became as bright as day. Then, over the building of the cinema "Zarya" appeared a ball lightning, ~1.5 m in diameter, having an intense orange color..... blazing briefly on the surface of the ground .... A strong explosion was heard.... The probability that the observed phenomena had been caused by ordinary linear lightning can be practically excluded."  It was estimated that ~ $1.1 \times 10^9$ J was released by the ball lightning in forming molten slag of 440 kg of ground and boiling 175 kg of water.

**15) Larger Activity Associated With Thunderstorms and Lightning**

Thunderstorm activity may be involved in the charging of little black holes, and/or the high fields associated with thunder clouds (Cobine, 1958) may attract charged little black holes. During lightning, runaway high energy charged particles in the high energy tail of the Maxwellian distribution have more of a chance of being ingested by the black hole due to their shorter de Broglie wavelengths. The potential of charged clouds may get as high as $10^9$ V (Rabinowitz, 1987). A startling recent discovery that γ-ray showers lasting from minutes to hours have been detected starting in the microseconds aftermath of some lightning flashes is reported by Krieger (2004). No conventional explanation has been found. To date, little black holes are as viable an explanation to account for this as any other.

# 11  Feasible Little Black Hole Geophysical and Astrophysical Processes

## 11.1  Overview

Many geophysical and astrophysical processes are not yet well understood. There would be profound implications if it could be established that LBH are the dark matter of the universe, and on rare occasion initiate tremors and trigger quakes. A testable LBH model of sporadic tremors and quakes is explored to determine under what conditions LBH may be relevant to geophysical and astrophysical processes.

Although the extraordinary weakness of gravity makes it by far the weakest of the interactions, viewing little black holes (LBH) as a class of elementary particles puts them in a league with hadrons as strongly interacting particles. They interact strongly both in the subatomic and macroscopic realms. The weakness of gravity is illustrated by the ratio of the gravitational force to the electric force of $2.4 \times 10^{-43}$ between two electrons (mass $9.1 \times 10^{-31}$ kg) and $8.0 \times 10^{-37}$ between two protons (mass $1.7 \times 10^{-27}$ kg). However two little black holes (LBH) each the size of a nucleon ($R_H \sim 10^{-15}$ m) have as much mass as a mountain ($10^{12}$ kg $\approx 10^9$ ton), completely turning this ratio around to $\sim 10^{41}$ which is well beyond normal strong interactions. The word "little" as used herein refers to the black hole radius, rather than its mass. As we shall see in Sec. 11.2.2, for very low mass LBH, the repulsive radiation force cannot be neglected.

Little black holes are expected to be made primarily, if not exclusively in the milieu of the high energies and high pressures of the big bang. This can be understood by looking at the extremely high density of LBH. To create a black hole (BH), an object of mass M must be crushed to a density

$$\rho = M / \left(\tfrac{4\pi}{3} R_H^3\right) = 7.3 \times 10^{79} M_{kg}^{-2} \text{ kg/m}^3, \text{ where} \tag{11.1}$$

$$R_H = 2GM/c^2 = 1.48 \times 10^{-27} M_{kg} \text{ m} \tag{11.2}$$

is the Schwarzchild radius, often also called the horizon of the BH. Thus a $10^{-3}$ kg LBH has $R_H \sim 10^{-30}$ m and $\rho \sim 10^{86}$ kg/m³ ($10^{83}$ g/cm³). A LBH the size of a nucleon ($R_H \sim 10^{-15}$ m) has a mass of $10^{12}$ kg and density $\rho \sim 10^{56}$ kg/m³ ($10^{53}$ g/cm³).

Here are some masses of familiar objects, and their corresponding radii and densities if they were compressed into being black holes.

$$M_{penny} \sim 1 \text{gm}, r \sim 10^{-28} \text{ cm}, \rho \sim 10^{83} \text{ gm/cm}^3$$
$$M_{mountain} \sim 10^9 \text{ton} \approx 10^{15} \text{gm}, r \sim 10^{-13} \text{ cm}, \rho \sim 10^{53} \text{ gm/cm}^3$$

$10^{-13}$ cm is the size of nucleons like protons and neutrons.

$$M_{earth} \approx 6 \times 10^{27} \text{gm}, r \sim 1 \text{cm}, \rho \sim 10^{27} \text{ gm/cm}^3$$
$$M_{sun} \approx 2 \times 10^{33} \text{gm}, r \sim 1\text{mi} \sim 10^5 \text{ cm}, \rho \sim 10^{16} \text{ gm/cm}^3$$

It is interesting to note that when an object about the mass of the sun becomes a black hole it is gravitationally crushed to $\sim$ nuclear density: $\rho \sim 10^{16}$ gm/cm³. By way of comparison ordinary heavy elements have

$$\rho_{Pb} = 11.3 \text{ gm/cm}^3, \rho_{Au,Pt,Os,Ir} \approx 20 \text{ gm/cm}^3.*$$

LBH can be characterized by a few variables such as mass, angular momentum, and electric charge just as is done with ordinary elementary particles. Nathan Rosen (1989 a, b) [of the Einstein, Podolsky, Rosen paradox] was one of the first scientists that considered a possible connection between elementary particles and LBH. LBH of Planck mass ($2.2 \times 10^{-8}$ kg and $10^{-33}$ m) with charges $\pm\frac{1}{3}e, \pm\frac{2}{3}e,$ and $\pm e$ were the starting point of his investigation.

In the numerical examples which follow, M ~ $10^{12}$ kg is used for illustrating the passage of a LBH through the earth, sun, and neutron stars. For levitating in the atmosphere, M <~ $10^{-3}$ kg, and as M decreases the LBH is repelled away from the earth, long before producing destructive radiation. In my model the LBH radiation is a function of the mass of the LBH, as well as both the distance and mass of the second body from the LBH; and is greatly attenuated relative to Hawking's. For Hawking it depends only on the LBH mass and would be $5.70 \times 10^9$ W for M = $10^{12}$ kg. For M = $10^{-3}$ kg, his would be $5.70 \times 10^{39}$ W, which is so exceedingly high that it could cause devastation.

**11.2 Gravitational tunneling radiation compared with extreme hawking radiation**

For an isolated LBH with M $\gtrsim$ $10 M_{Planck}$ = $2.2 \times 10^{-7}$ kg, the Hawking model predicts

$$P_{SH} = \left[\frac{\hbar c^6}{960\pi G^2}\right]\frac{1}{M^2} \sim 10^{47} \text{W}, \tag{11.3}$$

with a power density of $\dfrac{P_{SH}}{4\pi R_H^2} \sim 10^{105} \text{W/m}^2 = 10^{101} \text{W/cm}^2$. **This $10^{47}$ W may be the largest possible luminosity in nature from a single body**, if Hawking radiation were to exist. Even without Hawking radiation, this magnitude is possible from GTR in the highly unlikely limit as the tunneling probabiity approaches 1, for close encounters of LBH.

The Hawking high frequency luminosity of such an LBH is comparable to the visible luminosity of the entire universe (Rabinowitz, 2001 c):

$P_{univ}$ ~ (~$10^{26}$ W/star)(~$10^{12}$ stars/galaxy)(~$10^9$ to $10^{12}$ galaxies)

~$10^{47}$ to $10^{50}$ W. (11.4)

The argument cannot be made that there are presently no LBH with such a small mass. Even though in the Hawking model all LBH created in the big bang with M ≤ $10^{12}$ kg would have evaporated by now, originally more massive LBH can now have evaporated down to $\gtrsim$ $M_{Planck}$. But there has been no evidence for such extreme hawking radiation, even from larger LBH at lower luminosity but longer radiation duration. Such a glaringly large luminosity is not expected from Gravitational Tunneling Radiation $P_R$ as given by eq. (2.6) from Sec. 2:

$$P_R \approx \left[\frac{\hbar c^6 \langle e^{-2\Delta\gamma}\rangle}{16\pi G^2}\right]\frac{1}{M^2} \sim \frac{\langle e^{-2\Delta\gamma}\rangle}{M^2}\left[3.42 \times 10^{35} \text{W}\right]. \tag{2.6}$$

A very close encounter of two LBH would be required, and as we shall see, this is highly unlikely due to the beamed radiation between them that produces a repulsive force as shown in Sec. 11.3.2.

Another argument that favors $P_R$ is that the radiation is due to a tunneling process and not an information-voiding Planckian black body radiation distribution (Rabinowitz, 2003). Thus $P_R$ can carry information related to the formation of a BH, and avoid the information paradox associated with Hawking radiation. Also the reduced radiation of $P_R$ allows LBH to be candidates for the dark matter, i.e. 95% of the missing mass of the universe. For Hawking that many LBH would make toast of the universe. That is why he concludes that his LBH can't be more than one-millionth of the mass of the universe. Belinski (1995), a noted authority in the field of general relativity, unequivocally concludes "the effect [Hawking radiation] does not exist." Many significant consequences can result from a change in the model of black hole radiation.

## 11.3 Forces between neutral black holes

### 11.3.1 *Universal maximum attractive force*

The attractive force between two identical black holes (BH) of mass $M \gg M_{Planck}$ at a separation of $2BR_H$ (B > 1) is

$$F_A \approx \frac{GM^2}{r^2} = \frac{GM^2}{[2BR_H]^2} = \frac{GM^2}{\left[2B\left(\frac{2GM}{c^2}\right)\right]^2} = \frac{c^4}{16B^2G} \sim \frac{10^{43}\,N}{B^2} \quad . \tag{11.5}$$

This is a universal attractive force that acts between two identical black holes of any mass at all separations $\gg R_H$, provided that their separation is scaled in terms of the same multiple of $R_H$ (Rabinowitz, 2001 c). **For B = 1, F = $10^{43}$ N. which may be the largest possible attractive force in nature between two masses.** It is huge compared with any other force such as electrical, nuclear, etc. Using the largest force $F_A$, an even larger power $P = F_A \cdot c = c^5/16G \sim 10^{51}\,W$ than given by eq. (11.3) may be obtained. However, the idea here is to obtain the largest power and the largest forces in a physical context rather than just from a dimensionally proper combination of fundamental constants.

The expression (11.5) is only approximate since in my view the close proximity of two BH distorts the horizons on the adjoining sides of the BH. In addition, the Einsteinian effective potential of a BH is ~ four times stronger than the Newtonian potential near the BH, although the two are approximately equal for B > 10, i.e. for r > 10 $R_H$ (Rabinowitz, 1999c). Furthermore, eq. (11.5) neglects the radiative repulsive force due to the tunneling radiation between BH, which is discussed next.

### 11.3.2 *Universal maximum repulsive radiative force*

The repulsive radiative force between two black holes is

$$F_R \sim -c\left[\frac{dM}{dt}\right] \approx -c\left[\frac{-P_R}{c^2}\right] = \frac{P_R}{c} \approx \frac{1}{c}\left[\frac{\hbar c^6 \langle \Gamma \rangle}{16\pi G^2}\right]\frac{1}{M^2}, \qquad (11.6)$$

where $P_R$ is given by eq. (2.6) (in Sec. 2 and directly above), and the transmission coefficient $\Gamma \approx$ the tunneling probability $e^{-2\Delta\gamma}$ for LBH (Rabinowitz, 1999 a) since the emitted particle velocity $\approx c$ on both sides of the barrier. In the 0 angular momentum case with the origin at the center of mass of masses M and $M_2$:

$$\Delta\gamma = [\vec{b}_2 - \vec{b}_1]\left\{\frac{2\mu}{\hbar^2}\left[Gm\left(\frac{M}{r}+\frac{M_2}{r_2}\right)-E\right]\right\}^{\frac{1}{2}}, \qquad (11.7)$$

where $b_2$ and $b_1$ are the turning points of the potential barrier, the reduced mass $\mu = \frac{MM_2}{M+M_2}$, and the total energy $E = \frac{-GmM}{r-b_1} + \frac{-GmM_2}{r_2+b_1}$. For $M = M_2$, and primarily on-axis tunneling, eq. (3.3) reduces to

$$\Delta\gamma = [2b]\left\{\frac{M}{\hbar^2}\left[Gm\left(\frac{2M}{r}\right)-E\right]\right\}^{\frac{1}{2}}. \qquad (11.8)$$

The average tunneling mass is related to the BH mass through the BH temperature: $\langle m \rangle \approx kT/c^2 \propto 1/M$ (Rabinowitz, 1999 a). So it is not necessary to know in detail the nature of the emitted constituents.

Let us find the LBH mass for which the repulsive and attractive forces are comparable. As given by eqs. (11.5) and (11.6) $F_R \sim F_A$ yields

$$\frac{c^4}{16B^2 G} \sim \frac{1}{c}\left[\frac{\hbar c^6 \langle \Gamma \rangle}{16\pi G^2}\right]\frac{1}{M^2} \Rightarrow M \sim \left[\frac{\hbar c B^2 \langle \Gamma \rangle}{\pi G}\right]^{\frac{1}{2}}, \qquad (11.9)$$

For $B^2\langle\Gamma\rangle \rightarrow \sim 1$, eq. (10.9) yields $M \varnothing \sim M_{Planck}$, since $\langle\Gamma\rangle \xrightarrow[b\rightarrow\sim 0]{} \sim 1$ by eq. (11.8).

**Thus F = $10^{43}$ N may also be the largest possible repulsive force in nature between two masses.** (Rabinowitz, 2001 c)

In the very low probability <u>configurational</u> limit of $\langle\Gamma\rangle \sim 1$ [high tunneling probability], the form of $P_R$ looks like the Hawking radiated power $P_{SH}$, with an important distinction. $P_{SH}$ <u>is omnidirectional and does not yield repulsion in the standard Hawking model</u>, whereas $P_R$ is beamed between the two bodies resulting in repulsion. Two LBH must get quite close for maximum tunneling radiation. In this <u>configurational limit</u>, there is a similarity between the tunneling model and what may be expected from the Hawking model (1974, 1975), in that the tidal forces of two LBH would add together to give more radiation at their interface in Hawking's model. This should also produce a repulsive force, though somewhat smaller than from $P_R$, since there is also radiation in all directions.

One should use quantum gravity for such calculations, but it hasn't yet been formulated despite decades of dedicated work on this difficult subject. There may be concern regarding the use of semi-classical physics at the Planck scale of $\sim 10^{-35}$ m with energy $\sim 10^{19}$ GeV. However as measured at large distances, the gravitational red shift (Rabinowitz, 2003) substantially reduces the impact of high energies near LBH .

**11.4  LBH flux in the atmosphere**

For LBH coming to the earth from an extremely large distance in essentially free fall from the edge of the universe ($R_U \sim 1.4 \times 10^{26}$ m), by the conservation of energy we can calculate $v_{bh}$ their impact velocity at the earth assuming only free fall:

$$v_{bh} = \left[v_{LBH}^2 + \frac{2GM_e}{R_e} - \frac{2GM_e}{R_U}\right]^{1/2} \approx \left[v_{LBH}^2 + \frac{2GM_e}{R_e}\right]^{1/2}, \qquad (11.10)$$

where $M_e = 6.0 \times 10^{26}$ kg and $R_e = 6.4 \times 10^6$ m are the earth's mass and radius. If the initial LBH velocity $v_{LBH} = 0$, the impact velocity is $v_{bh} \sim 10^4$ m/sec. This is the rationale and LBH velocity used by others in the past. Note that by symmetry of eq. (11.10), the impact velocity equals the escape velocity.

However, a substantially larger velocity should be used. Since the LBH were created during the big bang, at a large distance from the earth they should be in the cosmic rest frame. The velocity of our local group of galaxies with respect to the microwave background, i.e. with respect to the cosmic rest frame (Turner and Tyson, 1999) is a reasonable velocity $v_{LBH} \sim 6.2 \times 10^5$ m/sec for the LBH with respect to the earth at $R_U$. Thus $v_{bh} \spadesuit v_{LBH}$. It is interesting that the free fall velocity from rest at $R_U$ to the sun is $6.2 \times 10^5$ m/sec $\approx v_{LBH}$, where $M_{sun} = 2.0 \times 10^{30}$ kg and $R_{sun} = 7.0 \times 10^8$ m are used in eq. (11.10). For a neutron star using $M_n = M_{sun} = 2.0 \times 10^{30}$ kg, and $R_n = 10^4$ m, eq. (11.10) gives $v_{bh} \spadesuit 1.6 \times 10^8$ m/sec $\spadesuit 0.5$ c, which is close to where relativistic effects become important. This should not be surprising as a neutron star is close to being a BH, where $v_{bh}$ would be c, the speed of light.

Massive LBH have reduced GTR and at large velocities do not slow down appreciably due to their large mass or angle of approach, and go right through the earth. (This interaction is covered in Sec. 11.4.) As will be shown, the LBH incidence rate matches the estimated BL rate well, but may be too low to account for much quake activity unless the heavier LBH are in re-entrant orbits.

In the Rabinowitz model (1999 a, b, c), those LBH that reach the earth's atmosphere and are small enough to have sufficient radiation reaction force to slow them down to the range of $10^{-2}$ to $10^2$ m/sec with a typical value $v_{BL} \sim 1$ m/sec, can manifest themselves as BL. In most cases $h/\tau << v_{BL}$. So Eq. (9.4) in Sec. 9 implies that the ball lightning current in the atmosphere ≈ the LBH current far away. We can thus give a range for the BL flux density as given by eq. (9.5) in Sec. 9:

$$\rho_L v_{LBH} < \rho_B v_{BL} < \rho_L v_{LBH}\left(\frac{A_{far}}{A_E}\right). \tag{9.5}$$

The distribution of LBH masses is not known. Assuming that LBH comprise all of the dark matter, i. e. 95 % of the mass of the universe (Rabinowitz, 1999 a) with 10% of the LBH average mass $\overline{M}_{LBH}$ ~ $10^{-3}$ kg which can linger in the atmosphere:

$$\rho_L \sim \frac{0.1(0.95 M_{univ} / \overline{M}_{LBH})}{V_{univ}}. \tag{11.11}$$

For $M_{univ}$ ~ $10^{53}$ kg and $V_{univ}$ ~ $10^{79}$ m$^3$ (radius of 15 x$10^9$ light-year = 1.4 x $10^{26}$ m), $\rho_L$ ~ $10^{55}$ LBH/ $10^{79}$ m$^3$ = $10^{-24}$ LBH/m$^3$. As shown in Sec. 9 my model predicts reasonable agreement with the references given there, that the incidence rate of BL is roughly in the range:

$$10^{-12} \text{ km}^{-2}\text{ sec}^{-1} \text{ to } >\sim 10^{-8} \text{ km}^{-2} \text{ sec}^{-1} \text{ for } A_{far}/A_E > \sim 10^4. \tag{11.12}$$

**11.5 Incidence rate of LBH through the Earth**

Assuming that 10% of the more massive LBH have an average mass $\overline{M}_{LBH}$ ~ $10^{12}$ kg, there are ~ $10^{40}$ LBH/ $10^{79}$ m$^3$ = $10^{-39}$ LBH/m$^3$, and with $(A_{far}/A_E)$ ~ $10^6$ eqs. (9.5) and (9.6) imply that the flux of these LBH through the earth is ~ $10^{-27}$/m$^2$-sec. Such LBH are too massive to produce enough exhaust radiation to linger in the atmosphere, and so go right through the earth. The earth's diameter is 1.3 x $10^7$ m which implies that the incidence rate of $10^{12}$ kg LBH is ~ ($10^{-13}$/sec) ~ 1 LBH /$10^5$ year. This could be augmented if the heavier LBH are in re-entrant orbits.

**11.6 Incidence rate of LBH through the sun**

With an average mass $\overline{M}_{LBH}$ ~ $10^{12}$ kg for 10% of the more massive LBH, gravitational focussing may increase their flux ~ $10^5$ times greater than through the earth to ~ $10^{-22}$/m$^2$-sec. The sun's diameter is 1.4 x $10^9$ m, implying an incidence rate of ~ $10^{-9}$/sec ~ 1 LBH /10 year through the sun, neglecting LBH re-entrant orbits.

**11.7 Incidence rate of LBH through neutron stars**

With a gravitational enhancement ~ $10^{10}$ with respect to the earth, the flux is ~ $10^{-17}$/m$^2$-sec for the heavier LBH through neutron stars. The typical diameter of neutron stars of $\sim 10^4$ m, implies an incidence rate of $\sim 10^{-9}/\text{sec} \sim 1$ LBH /10 year through neutron stars, neglecting re-entrant orbits for the LBH.

## 12 Little Black Hole Transmission Through Matter

### 12.1 LBH gravitational impulse transfer

The change in momentum of a particle of mass m of negligible initial velocity due to the impulse imparted by a LBH of mass M and velocity $v_{bh}$ as it passes by is

$$m\Delta v = (F)\Delta t = \left(\frac{GMm}{b^2}\right)\frac{2b}{v_{bh}} = \frac{2GMm}{bv_{bh}} \approx mv_f, \qquad (12.1)$$

where $v_f$ is the particle's final velocity directed radially inward toward the center line of the LBH trajectory, and b is the impact parameter. The energy lost by the LBH equals the energy gained by m

$$\Delta E = \tfrac{1}{2}mv_f^2 = \tfrac{1}{2}m\left(\frac{2GM}{bv_{bh}}\right)^2 = \frac{2G^2M^2m}{b^2v_{bh}^2}, \qquad (12.2)$$

In this section m represents the target constituents such as the mass of a neutron, a typical air molecule ($N_2$), or a typical rock molecule ($SiO_4$).

We can make a rough estimate of the maximum temperature that the LBH can produce in its wake, assuming that all the energy is converted into heat with negligible heat conduction, and neglecting heats of vaporization and fusion (heat of vaporization >> heat of fusion). Thus $\Delta E \approx \tfrac{3}{2}kT_{max}$ in eq. (12.2) implies

$$T_{max} \approx \frac{4G^2M^2m}{3kb^2v_{bh}^2}. \qquad (12.3)$$

Depending on the magnitude of the different variables, it is possible to exceed the melting point of rock $\sim 1500$ °C. The actual temperature can be much less depending on how much of the energy is partitioned into a shock wave. This would depend on the nature of the part of the earth traversed (e.g. rock, liquid, etc.) and on the magnitude of the energy loss and power input per atom. Thus energy is partitioned differently into heat, tremor, and shock wave.

The LBH energy loss per unit length is

$$\frac{dE}{dx} = \int N\Delta E\left(\frac{2\pi nb\,db}{N}\right) = \int \frac{2G^2M^2m}{b^2v_{bh}^2}(2\pi nb\,db) = \frac{4\pi G^2M^2mn}{v_{bh}^2}\int_{b_{min}}^{b_{max}}\frac{db}{b}$$
$$= \frac{4\pi G^2M^2\rho}{v_{bh}^2}\ell n\left(\frac{b_{max}}{b_{min}}\right) \qquad (12.4)$$

where N is the number of target particles, $n = N/2\pi b\,db\,dx$ is the number density of the target particles, and $\rho = mn$ is the mass density of these particles.

From eq. (12.4), the total power dissipated by each LBH is thus

$$P = \frac{dE_{total}}{dt} = \frac{NdE}{dx/v_{bh}} = \frac{4\pi G^2 M^2 \rho}{v_{bh}} \ln\left(\frac{b_{max}}{b_{min}}\right). \quad (12.5)$$

If $\Delta E$ = ionization potential $V_i$ of an atom of mass m, substituting $V_i$ into eq. (12.2) yields the ionization parameter

$$b_i = \left[\frac{2G^2 M^2 m}{V_i v_{bh}^2}\right]^{1/2} = \frac{GM}{v_{bh}}\left[\frac{2m}{V_i}\right]^{1/2}. \quad (12.6)$$

The maximum impact parameter $b_{max} \geq b_i$. Let us next examine a case when the > sign applies.

**12.2 Gravitationally enhanced ionization parameter**

The intense gravitational field of a LBH causes more atoms to be ionized than given by only kinetic considerations since atoms will be gravitationally captured in orbit around the LBH with the ultimate fate of being ionized even if they do not fall into the BH. This clearly occurs for free particles in the atmosphere, and may also occur if matter is temporarily vaporized along the path of a LBH going through the earth. As shown by eq. (6.1) of Sec. 6, the gravitational potential energy of a particle of mass m in the field of a LBH is

$$V = -\frac{GMm}{r} - p\left(\frac{GM}{r^2}\right) - \frac{\alpha_p}{2}\left(\frac{GM}{r^2}\right)^2, \quad (6.1)$$

where p is the permanent dipole moment, and $\alpha_p$ is the gravitational polarizability. This led to an effective ionization radius eq. (6.5), which written more explicitly is:

$$r_E = b_i\left[1 - \frac{1}{\frac{3}{2}kT}\left(-\frac{GMm}{b_i} - p\left(\frac{GM}{b_i^2}\right) - \frac{\alpha_p}{2}\left(\frac{GM}{b_i^2}\right)^2\right)\right]^{1/2}. \quad (12.7)$$

This is greater than the ionization parameter $b_i$ given by eq. (12.6) so that $b_{max} = r_E \geq b_i$ for a gaseous medium.

If the medium is not gaseous or does not become vaporized, then according to Greenstein and Burns (1984):

$$b_{max} = b_{sonic} = \frac{2GM}{v_{bh} c_s}, \quad (12.8)$$

where $c_s$ is the speed of sound in the medium.

**12.3 Minimum impact parameter**

The minimum impact parameter is determined by quantum mechanics since quantum effects smear out the particle and reduce the probability of its ingestion in the LBH. Though different approaches agree that this occurs when the particle is absorbed by the LBH, they give substantially different values. Fortunately this does not make a big difference in the LBH energy loss per unit length nor in the total power dissipated per LBH as given by eqs. (12.4) and (12.5) since these have a weak logarithmic dependence $\ln(b_{max}/b_{min})$.

From one point of view a target particle cannot be absorbed in a black hole unless its Compton wavelength ≤ the LBH Schwarzchild (horizon) radius $R_H$; and that it has a very low probability of being absorbed unless its de Broglie wavelength ≤ $R_H$. Thus
$$b_{min} \sim \lambda = \frac{h}{mv_{bh}} = \frac{2GM}{c^2}, \qquad (12.9)$$
where the relative velocity between the approximately stationary particle and the LBH is the velocity $v_{bh}$ of the LBH. Equation (12.9) would apply provided that $\lambda$ is the shortest length scale in the LBH rest frame.

Another criterion for absorption applies only to very small LBH. It is that $\lambda$ does not change appreciably in a length scale comparable to itself. Interestingly, this implies that $v_{bh} \sim 2GMm/h < c$. Thus for m ~ atomic mass, only LBH with M < $10^{12}$ kg ($R_H$ =$10^{-15}$ m) can absorb atoms. This criterion would not apply for much larger BH. If this criterion is correct, then even if the classical orbital radius of the particle were small enough to allow it, the particle Compton wavelength for absorption would need to be less than the LBH radius for all BH. We can essentially set $b_{min}$ = Compton wavelength:
$$b_{min} \sim \lambda_C = \frac{h}{mc}. \qquad (12.10)$$
A classical orbital approach (Zeldovich and Novikov, 1971) using the Einsteinian effective potential of the LBH which is ~ four times stronger than the Newtonian potential near the LBH yields
$$b_{min} \sim \frac{4GM}{cv_{bh}} \qquad (12.11)$$
for ingestion and is independent of m as would be expected from the equivalence principle.

## 13 LBH Energy Loss, Power Dissipation, and Range

### 13.1 LBH orbits unlikely inside earth, sun, and neutron stars

For a closed or quasi-closed (non re-entrant) circular orbit of radius r inside a body of mass density ρ:
$$\frac{Mv_{bh}^2}{r} = \frac{GM[\rho \frac{4}{3}\pi r^3]}{r^2} \Rightarrow r = \frac{v_{bh}}{[\frac{4}{3}\pi G\rho]^{1/2}}. \qquad (13.1)$$

-31-

Equation (12.1) indicates that such orbits execute simple harmonic motion with constant angular velocity for constant $\rho$, since $\omega = v_{bh}/r = [\frac{4}{3}\pi G\rho]^{1/2}$ = constant.

1) For the earth with an average density of $\rho = 5.5 \times 10^3$ kg/m$^3$, and $v_{bh} = 6.2 \times 10^5$ m/sec, r = $5.0 \times 10^8$ m >> $R_e = 6.4 \times 10^6$ m. So an internal orbit inside the earth is not possible unless the LBH velocity is greatly reduced.

2) For the sun with an average density of $\rho = 1.4 \times 10^3$ kg/m$^3$, and $v_{bh} = 8.7 \times 10^5$ m/sec, r = $1.2 \times 10^9$ m > $R_s = 7 \times 10^8$ m. So an internal orbit near the limb of the sun would almost be possible.

3) For a neutron star with an average density of $\rho = 4.8 \times 10^{17}$ kg/m$^3$, and $v_{bh} = 1.6 \times 10^8$ m/sec, r = $1.4 \times 10^4$ m > $R_n = 10^4$ m. So an internal orbit near the limb of the neutron star would almost be possible.

As we shall see in the next sections, velocity degradation of LBH is difficult to achieve by ordinary collisional-like interactions because dE/dx is relatively small. It is also difficult to reduce the LBH velocity by particle absorption, since LBH particle absorption is a very low probability event, and when it does occur for particle mass m << M, there is hardly any decrease in $v_{bh}$.

**13.2 LBH interaction in going through the earth**

From eqs. (12.6) and (12.7), with an ionization potential ♠ 15 eV = $2.4 \times 10^{-18}$ J, and a LBH velocity of $6.2 \times 10^5$ m/sec, $10^{12}$ kg LBH have an upper limit $b_{max} = r_E \sim 5 \times 10^{-4}$ m. If $b_{sonic} ♠ b_{ioniz}$, then $b_{max} \sim 10^{-8}$ m as given by eqs. (12.8) and (12.6). The minimum impact parameter $b_{min} \sim 10^{-17}$ m, or $10^{-15}$ m, or $10^{-12}$ m, as given by eqs. (12.9), or (12.10), or (12.11). Because of the logarithmic dependence it does not make much difference which of these $b_{max}$ or $b_{min}$ is used. Thus by eq. (12.4) dE/dx ~ $10^{-2}$ J/m in going through the earth. The overall density of the earth is $5.5 \times 10^3$ kg/m$^3$ (5.5 gm/cm$^3$). The mantle density (first 50 miles in from the surface) is 2.7 gm/cm$^3$.

From eq. (12.4), the power dissipated per LBH is

$$P = v_{bh}\frac{dE}{dx} = \frac{4\pi G^2 M^2 \rho}{v_{bh} N} \ell n\left(\frac{b_{max}}{b_{min}}\right). \qquad (13.2)$$

Thus P ~ $10^4$ W/LBH for M ~ $10^{12}$ kg. (This is small compared to the total power output of $4.2 \times 10^{13}$ W emanating from inside the earth (Stacey, 1992). From eq. (12.4), the total energy input to the earth per such LBH is

$$E_t \sim \frac{dE}{dx}(\sim R_e) = (10^{-2} J/m)6.4\times10^6 m \sim 10^5 J/LBH. \qquad (13.3)$$

This is an insignificant energy loss for a LBH with incident velocity of 6.2 x $10^5$ m/sec and kinetic energy of 2 x $10^{23}$ J.

The range of a LBH

$$\mathfrak{R} = \frac{E}{dE/dx} = \frac{\frac{1}{2}Mv_{bh}^2}{\left[\frac{4\pi G^2 M^2 \rho}{v_{bh}^2}\right]\ln\left(\frac{b_{max}}{b_{min}}\right)} = \frac{v_{bh}^4}{8\pi G^2 M \rho \ln\left(\frac{b_{max}}{b_{min}}\right)}. \quad (13.4)$$

Equation (13.4) implies that $E = E_o e^{-x/\mathfrak{R}}$. So the range is that path length when the LBH energy has fallen to 1/e of its initial value, $E_o$. Neglecting black hole decay, the range would be 3 x $10^{25}$ m through solid earth of density 5.5 x $10^3$ kg/$m^3$, which is 21% of the radius of the universe ($R_U$ ~ 1.4 x $10^{26}$ m). To put this into perspective, if the earth had a radius $R_E$ ~ 6 x $10^8$ m (100 times larger than its actual radius), then a LBH with $v_{bh}$ = 6.2 x $10^5$ m/sec in circular orbit just inside this larger earth would make ~ $10^{16}$ revolutions in 1500 billion years i.e. ~100 times longer than the present age of the universe. So orbits that are re-entrant into the core of the earth (as well as the sun and neutron stars), could easily persist for almost endless cycles.

### 13.3 LBH interaction in going through the sun

Equation (12.4), for an upper limit using the sun's core density of ~ $10^5$ kg/$m^3$ and $\ln(b_{max}/b_{min})$~30, yields dE/dx ~ $10^{-1}$ J/m for a LBH of M ~ $10^{12}$ kg. From eq. (10.2), the maximum power dissipated in the sun is P ~ $10^5$ W/LBH. Even at this high density the LBH range would be 5 x $10^{25}$ m neglecting black hole decay, which is 38 % of the radius of the universe. If the sun's radius were 1.2 x $10^9$ m (almost a factor of 2 larger than the actual radius $R_s$ = 7 x $10^8$ m), then a LBH with $v_{bh}$ = 6.2 x $10^5$ m/sec could be in a circular orbit just inside such a larger sun. It would make ~ $10^{16}$ revolutions in 2000 billion years i.e. ~130 times longer than the present 13.7 billion-year age of the universe.

A startling conjecture was made by Hawking (1971) to account for the missing solar neutrino flux (Kim et al, 1993) that a black hole has fallen into the center of the sun and is gobbling up neutrinos. If such a black hole did not first evaporate away by Hawking radiation, it would eventually cause the sun to collapse.

### 13.4 LBH interaction in going through neutron stars

The gravitational potential energy of a neutron star is

$$V \sim \frac{GM_n^2}{R_n}, \quad (13.5)$$

where we will take the neutron star mass $M_n$ ~ solar mass = $2 \times 10^{30}$ kg, with a radius $R_n$ ~ $10^4$ m. This yields a potential energy of $10^{46}$ J. The binding energy of a neutron of mass $m_n = 1.67 \times 10^{-27}$ kg is

$$\Delta E_n \sim \frac{GM_n m_n}{R_n} = \frac{GM_n^2}{R_n M_n / m_n} = \frac{10^{46} \text{ J}}{M_n / m_n} \approx 2.2 \times 10^{-11} \text{ J/neutron}. \quad (13.6)$$

This is $10^2$ MeV which is quite large even compared with nuclear binding energies of 6 - 8 MeV/nucleon. If in eq. (12.6), we set $V_i = \Delta E_n \sim 10^{-11}$ J, we obtain $b_{max} \sim 10^{-13}$ m. In this view, a LBH can displace a neutron by gravitational interaction as it goes through a neutron star.

From another point of view, we may think of the interaction of a LBH with neutrons as analogous to the interaction of a LBH in ionizing an atom. A free neutron decays into a proton + electron + antineutrino with a half-life of 10.6 minutes. We may think of the ionization potential of a neutron as $< \sim m_n c^2 - m_p c^2$, the energy difference between the neutron and the proton. Thus $\Delta E_n \sim$ 939.56 MeV - 938.27 MeV = 1.29 MeV. In this scenario, eq.(12.6) yields $b_{max} \sim 10^{-12}$ m.

In the latter scenario and to some degree in the former, $\ell n(b_{max}/b_{min}) \sim 10$, and eq. (12.4) yields $dE/dx \sim 10^4$ J/m for a LBH of M ~ $10^{12}$ kg, and a neutron star density of $5 \times 10^{14}$ kg/m$^3$. The total energy lost is $10^4$ J/m (~$10^4$ m ) ~ $10^8$ J per LBH. The power dissipated is $1.6 \times 10^8$ m/sec ($10^4$ J/m) ~ $10^{12}$ W.

From eq. (13.4), neglecting black hole decay, the range would be ~ $10^{24}$ m through a neutron star, which is 1% of the radius of the universe. For an orbit just inside a neutron star with $R_n \approx 1.4 \times 10^4$ m, a LBH with $v_{bh} = 1.6 \times 10^8$ m/sec would make ~ $10^{19}$ revolutions in 0.25 billion year i.e. 1.6 % of the present age of the universe.

## 14 Devastation of Tungus, Siberia

The devastation of the Tungus region of central Siberia on June 30, 1908, remains a mystery to this day, despite the fact that there were large numbers of eyewitnesses and we know precisely when and where this gigantic explosion took place. A brilliant ball of fire crossed the sky and exploded in the atmosphere with a blast equivalent ~ $10^{15}$ to $10^{17}$ J (~30 million tons of TNT) (Krinov, 1966). More cataclysmic than a hydrogen bomb, the force flattened trees causing them to point radially outward within a 40-mile diameter circle; and hurled creatures like horses to the ground more than 400 miles from Tungus in the area of Kansk.

One of the many speculations that have been considered over the years is that this destruction of an area of more than 1200 square miles was caused by a ~$10^{17}$ kg LBH of atomic radius $10^{-10}$ m (Jackson and Ryan, 1973). Hawking (1971) proposed that the Tungus event resulted from the passage of a small black hole through the earth.

Burns et al. (1976) conclude that "$10^{28}$ to $10^{30}$ erg [$10^{21}$ to $10^{23}$ J] of seismic energy would have been deposited in the Earth... ". This is not only tremendously greater than actually recorded, but greater than some of the largest earthquakes (~ $10^{17}$ J) ever recorded. One of the biggest, the 1960 Chilean earthquake, released > ~ 4 x $10^{17}$ J, which is large compared with the average annual seismic energy release of 5 x $10^{17}$ J/yr = 1.5 x $10^{10}$ W (Stacey, 1992). Burns et al concluded that the Tungus catastrophe could not have been caused by an LBH. Although Greenstein and Burns (1984) included additional energy release due to Hawking radiation in a later paper (in Fig. 1 of their paper in which the $b_{ion}$ scale appears to be low) not related to the Tungus event, this was not included in their papers on Tungus. These papers did not consider that LBH might be the missing mass of the universe, which is a distinct possibility in the GTR model of LBH radiation as discussed in this Chapter.

Although the conclusion of Burns et al (1976) may well be correct, one may get much smaller numbers for the total energy released by an LBH going through the earth in Siberia. From eq. (13.2) the energy release for the example $10^{12}$ kg LBH at 6.2 x $10^5$ m/sec in this paper is only $10^5$ J with a correspondingly larger quadratic effect for larger M LBH. This large disparity results from the scaling of the energy input, where neglecting the logarithmic dependence, $E_t \propto (M/v_{bh})^2$. They assigned the impact velocity $v_{bh}$ ~ $10^4$ m/sec to their $10^{17}$ kg LBH. Jackson and Ryan (1973) used a similarly low $v_{bh}$ for their $10^{17}$ to $10^{19}$ kg LBH in concluding that, "total energy in the blast wave would be $10^{22}$ to $10^{24}$ erg [$10^{15}$ to $10^{17}$ J]."

As to the large energy release in the atmosphere there are other possibilities besides the impulse energy transfer considered. These include a charged LBH, and (even without Hawking radiation) the explosive disruption of the rotational energy outside an LBH because conservation of angular momentum prevents outside matter from falling into the LBH.

Accordingly this intriguing question may not have been decided so conclusively as yet. An LBH may still be ingesting part of Siberia since LBH take millions of years to consume objects that are considerably more voluminous than themselves. However it is more likely that if it were an LBH, the LBH went through the earth and exited. Neglecting re-entrant orbits, it is unlikely that a heavy LBH > $10^{12}$ kg will come again before more than $10^5$ years as indicated by my calculations in Sec. 9 combined with Sec. 11.4. So humankind may not have to worry about such a horrific natural occurrence happening for a long time.

## 15 Change in Angular Momentum Due to LBH Interaction

### 15.1 General

The total vector sum of the angular momentum of an incident LBH, $L_{bh}$ plus the spin angular momentum S of the target body is conserved because there is no external torque. The final velocity of an LBH as it emerges after travelling a distance r through the target body is

$$v_f = v_{bh}\left[1 - \frac{2\Delta E}{Mv_{bh}^2}\right]^{1/2} \approx v_{bh}\left[1 - 2\frac{2\Delta E}{Mv_{bh}^2}\right] = v_{bh} - \frac{4\Delta E}{Mv_{bh}}, \quad (15.1)$$

where $\Delta E \sim (dE/dx)(\sim r)$, and it was shown above that the second term in the square root factor is << 1. Thus the decrease in the magnitude of the LBH angular momentum is

$$\Delta L_{bh} = \vec{d} \times M(\vec{v}_{bh} - \vec{v}_f) \approx dM\frac{4\Delta E}{Mv_{bh}} \sim \frac{4d(rdE/dx)}{v_{bh}}, \quad (15.2)$$

where d is the moment arm with respect to the center of mass of the target body. The initial spin angular momentum of the target body, $S = \frac{2}{5}M_tR_t\omega_o^2$, where $\omega_o$ is its initial angular velocity. By conservation of the total angular momentum of the system, $\Delta \vec{S} = -\Delta \vec{L}_{bh}$.

**15.2 Neutron Star Pulsars**

Neutron star pulsars emit pulsed radiation that range from x-ray to radio frequencies (Davies,1992). The detection of polarization of the radiation, and of the rotation of the plane of polarization within a pulse was an indication that a strong magnetic field plays an important role in the pulses as the neutron star rotates, much like a lighthouse beacon produces a pulse of light in a given direction. The general tendency of pulsars to slow down, as well as the cyclotron radiation signature of x-ray pulsars has been explained in terms of huge magnetic fields ~ $10^6$ to $10^9$ Tesla ($10^{10}$ to $10^{13}$ G). It is theorized that when a star like the sun collapses rapidly with an initial magnetic field of $10^{-2}$ T, the field gets compressed due to high conductivity followed by a state of extremely high temperature superconductivity, which leads to

$$B_{final} = B_o\left[r_o^2/r_f^2\right] \sim 10^{-2} \text{ T } [(10^9\text{m})^2/(10^4\text{m})^2] \sim 10^8 \text{ Tesla}. \quad (15.3)$$

For a $10^{12}$ kg LBH going through a neutron star, with d ~ 0.5 $R_n$ ~ 0.5 x $10^4$ m, r ~ $R_n$, by eq. (15.2) $\Delta S_n = -\Delta L_{bh} \sim 10^4$ kg-m²/sec. This is a relatively small change in angular momentum of the neutron star that could be much larger for a more massive LBH since $\Delta S_n = -\Delta L_{bh} \propto M^2$. Two things may be of interest. One is that this impulse is imparted in a relatively short time ~ $R_n/v_{bh}$ ~ $10^4$ m/1.6 x $10^8$ m/sec ~ $10^{-4}$ sec. This is short compared with the 1 to >10 msec period of pulsars. The second is that this can lead to an increase in the pulsar frequency about half of the time, since roughly half the time the spin will increase and half the time the spin will decrease. The frequency increase may be significant as next discussed, but leaves open the question of observation of a decrease.

Dissipative mechanisms lead to a gradual slowdown of all radio pulsars. However occasionally there is a sudden increase in frequency, called a starquake, followed by another moderate slowdown of the frequency. This process is recurrent.

The slowdown time period for most pulsars is $10^3$ to $10^7$ years. It is thought that the abrupt frequency increase is related to a breakup of the surface crust leading to a decrease in moment of inertia (Shapiro and Teukolsky, 1983). However it is not clear that such a rearrangement including achievement of a new equilibrium position can occur rapidly enough, and it cannot account for the precipitous frequency increases observed in the Vela pulsar because they occur too frequently (Davies, 1992).

The latter is also a problem for my LBH spin-up mechanism unless the LBH are in re-entrant orbits which go in and out of the neutron star. In a mechanism that is similar in spirit to this, Stephen Hawking (1971) suggested that a $10^{14}$ kg LBH at the center of a neutron star "would produce a slight shrinking of the surface and might possibly be the cause of the recently observed pulsarquakes." For his mechanism it is also not clear that this process could produce a sufficiently rapid (for most pulsars) and frequent (such as Vela) decrease in the pulsar moment of inertia. Also both the Shapiro and Teukolsky, and Hawking deformation mechanisms should lead to power dissipation of the superconducting currents that maintain the high magnetic field (Rabinowitz, 1970, 1971).

**15.3  Earth and Sun**

Such effects do not appear so clearly on the earth and on the sun because changes in the rotational frequency are not as precisely and dramatically observed. For the earth a $10^{12}$ kg LBH produces $\Delta S_e = -\Delta L_{bh} \sim 10^5$ kg-m$^2$/sec. by eq.(15.2); and for the sun $\Delta S_s = -\Delta L_{bh} \sim 10^{10}$ kg-m$^2$/sec. Both are small compared with the spin angular momenta of the earth and sun. With a spin period of 1 day, the earth has $S_e = 7.2 \times 10^{33}$ kg-m$^2$/sec. For the sun's spin period of 24.7 day, the sun has $S_S = 1.2 \times 10^{42}$ kg-m$^2$/sec.

**15.4  Discussion of LBH interactions**

The enormous force between BH at close distances as scaled by $R_H$ is surprisingly matched by a comparably large repulsive radiative force as shown in Secs. 11.3.1 and 11.3.2. Whether massive compact objects (herein called LBH) are indeed black holes or Yilmaz (1958, 1982) gray holes should leave much of the analysis of this paper unchanged, but would shatter the Hawking model (1974, 1975) which requires a black hole horizon. The Rabinowitz model of LBH radiation in avoiding the unreasonably high radiation of Hawking, permits LBH to be considered as candidates for dark matter in the universe, and ball lightning on earth. The incidence rate of low mass LBH agrees well with the incidence rate of ball lightning.

Sections 12 and 13 have demonstrated that heavy LBH are unlikely initiators of seismic activity in the earth in terms of frequency of occurrence, unless there is a concentration mechanism such as re-entrant orbits. In terms of magnitude, it would be possible for <u>very heavy</u> LBH to contribute to seismic activity in the earth and in neutron stars directly or by triggering metastable sites. Section 15 indicated that LBH are capable of causing abrupt pulsar frequency changes. A LBH of mass M ~ $10^{12}$ kg was used in these example calculations. Since the energy deposition scales roughly as $M^2$, a heavier LBH can have a correspondingly bigger quadratic effect. Although it is unlikely that a LBH was responsible for the catastrophic 1908 event in the Tungus region of Siberia, analysis in this paper indicates that past conclusions ruling out a LBH may not be on as firm a basis as formerly thought because important input was overlooked.

## 16  Viable Black Hole Atoms

### 16.1  Black holes are ideal for making gravitationally bound atoms

Ordinary gravitational orbits, such as planetary orbits, are in the high quantum number, continuum classical limit. In considering gravitationally bound atoms (GBA), black holes are ideal candidates for the observation of quantization effects (Rabinowitz, 1990, 2001 a,b), since for small orbits very high density matter is necessary. Furthermore, "A little black hole can trap charge internally and/or externally. It could easily trap ~ 10 positive or negative charges externally and form a neutral or charged super-heavy atom-like structure (Rabinowitz. 1999a)." Moderately charged black holes could form electrostatically and gravitationally bound atoms. For the present let us consider only gravitational binding where the black hole mass M >> m, the orbiting mass. To avoid complications related to quantum gravity, m can be considered to be made of ordinary matter such as a nucleon or group of bound nucleons. We will also avoid the complication of the interaction of the orbiting body with GTR from the black hole.

Newtonian gravity is generally valid for r > 10 $R_H$ since the difference between Einstein's general relativity and Newtonian gravitation gets small in this region. (The black hole horizon or Schwarzschild, radius is $R_H = 2GM/c^2$, where M is the mass of a black hole and c is the speed of light.) This approximation should be classically valid for all scales since the |potential energy|:

$$|V| = \frac{G(M)\gamma m}{r} < \frac{G\left(R_H c^2/2G\right)m}{10 R_H} = \frac{\gamma m c^2}{20}, \qquad (16.1)$$

is scale independent, where $\gamma = \left(1 - v^2/c^2\right)^{-1/2}$. Thus it is necessary that |V| be smaller than 1/20 of the rest energy of the orbiting body of mass m.

### 16.2  Black hole atoms in the realm of Newtonian gravity

We will operate in the realm of Newtonian gravity and thus require the orbital radius r > 10 $R_H$. From (Rabinowitz, 2003) eq. (4.5) for 3-space i.e. n = 3, with principal quantum number j =1, and M >> m:

$$r_n = \left[ \frac{j\hbar \pi^{\frac{n-2}{4}}}{m[2G_n M \Gamma(n/2)]^{1/2}} \right]^{\frac{2}{4-n}} . \tag{16.2}$$

$$= \frac{\hbar^2}{(GM)m^2} = \frac{\hbar^2}{(R_H c^2/2)m^2} \geq 10 R_H$$

Solving eq. (16.2)

$$R_H \leq \frac{\hbar}{\sqrt{5} mc} = \frac{\lambda_C}{\sqrt{5}}, \tag{16.3}$$

where $\lambda_C$ is the reduced Compton wavelength of the orbiting particle. So $r \geq 10\, R_H$ is equivalent to the quantum mechanical requirement $\lambda_C \geq \sqrt{5} R_H$.

Now let us determine a relationship between M and m that satisfies $r \geq 10\, R_H$.

$$r = \frac{\hbar^2}{GMm^2} \geq 10 R_H = 10\frac{2GM}{c^2} . \tag{16.4}$$

Equation (16.4) implies that

$$Mm \leq \frac{\hbar c}{\sqrt{20 G}} = \frac{(M_{Planck})^2}{\sqrt{20}} . \tag{16.5}$$

For $M = m$, r is a factor of 2 larger from eq. (16.2), and eqs. (16.4) and (16.5) would yield $M \leq M_{Planck}/\sqrt{10}$. This is why it would be impossible to also avoid the realm of quantum gravity if the two masses are equal as is primarily done in Chavda and Chavda (2002).

We can now determine the ground state orbital velocity v in general for any M and m that satisfy $r \geq 10\, R_H$. By substituting eq. (16.5) into eq. (4.6) for v from (Rabinowitz, 2003) for $n = 3$, with $j = 1$ and $M \gg m$:

$$v_n = \left\{ \left[ \frac{2\pi G_n M \Gamma\left(\frac{n}{2}\right)}{\pi^{n/2}} \right] \left[ \frac{m^{2/(4-n)} \left[ 2G_n M \Gamma\left(\frac{n}{2}\right) \right]^{\frac{1}{4-n}}}{(j\hbar)^{2/(4-n)} \pi^{(n-2)/2(4-n)}} \right] \right\}^{1/2} . \tag{16.6}$$

$$= \frac{G(Mm)}{\hbar} \leq \frac{G}{\hbar}\left( \frac{\hbar c}{\sqrt{20 G}} \right) = \frac{c}{\sqrt{20}} = 0.224 c$$

So special relativity corrections would only be small here. However, not so in Chavda and Chavda (2002) where in some cases they have $v \approx c$.

Substituting eq. (16.6) for v into (Rabinowitz, 2003) eq. (4.6), the binding energy is

$$E = -\frac{m}{2}\left[ \frac{GMm}{\hbar^2} \right]^2 = -\frac{m}{2}[v^2] = -\frac{m}{2}\left[ \frac{c}{\sqrt{20}} \right]^2 = -\frac{mc^2}{40} . \tag{16.7}$$

A large range of M >> m can satisfy these equations. For a numerical example, let m = $m_{proton}$ = 1.67 x $10^{-27}$ kg. Equation (16.5) implies that M = 6.36 x $10^{10}$ kg, with $R_H$= 9.43 x $10^{-17}$m. Equation (16.7) gives a binding energy E = 3.76 x $10^{12}$ J = 23.5 MeV, with v = 6.72 x $10^7$ m/sec. We want the binding energy E >> kT, so T must be << 2.72 x $10^{11}$K. Since this is much less than the unification temperature $T_{unif}$ ~ $10^{29}$ K, such atoms would not be stable in the very early universe. However if they were formed at later times, they could be stable over most of the age of the universe, assuming negligible Hawking radiation (Rabinowitz, 1999 a and 2001 a, b).

**16. 3 Gravitational fine structure or coupling constant**

Just as the coupling in an ordinary electrostatic atom can be characterized by the electromagnetic fine structure constant, $\alpha$, we can also characterize the coupling in a gravitational atom by a gravitational fine structure constant, $\alpha_G$. However the designation of what $\alpha_G$ should be, is not nearly as clear as the designation of $\alpha$. The fine structure constant, $\alpha$, is a dimensionless coupling constant that characterizes the relatively weak electromagnetic interaction. The measure of the strength of the interaction between any two particles is given in general by the dimensionless ratio:

$$\frac{\text{[mutual force acting on the particles]} \times \text{[square of the distance between them]}}{\hbar c} . \quad (16.8)$$

This measure is called the "coupling constant" for the particles concerned.

Using SI units, the electrostatic force acting on each of two particles of charge -e and +e spaced distance r apart is

$$F = -\frac{e^2}{4\pi\varepsilon r^2}, \quad (16.9)$$

where $\varepsilon$ is the permittivity of free space. Thus in accord with eq. (16.8):

$$|\alpha| \equiv \frac{\left(\frac{e^2}{4\pi\varepsilon r^2}\right)}{\hbar c} r^2 = \frac{e^2}{4\pi\varepsilon \hbar c} \approx 1/137, \quad (16.10)$$

is the fine structure constant introduced by Arnold Sommerfeld (1916) in connection with his explanation of the fine structure of atomic spectra such as for hydrogen.

Traditionally, the gravitational fine structure (coupling) constant is given by

$$\alpha_G = \frac{G m_p^2}{\hbar c} \approx 5.88 \times 10^{-39}, \quad (16.11)$$

where $m_p$ = 1.67 x $10^{-27}$ kg is the mass of the proton. This is an extremely small coupling constant. However, the choice of the proton mass is arbitrary since no one has observed a gravitational atom -- much less one with two proton masses orbiting around each other.

In general for any two masses M and m,

$$\alpha_G = \frac{\left(\frac{GMm}{r^2}\right)}{\hbar c} r^2 = \frac{GMm}{\hbar c}. \tag{16.12}$$

Now in 3-space (n = 3) from eq. (16.6) we find for the ground state (j = 1) velocity of the orbiting mass m

$$v = \frac{GMm}{\hbar} = \alpha_G c. \tag{16,13}$$

This is the direct analog of the ground state velocity of the orbiting electron in a hydrogen atom

$$v = \alpha c \approx c/137. \tag{16.14}$$

If the two masses are each the Planck mass, $m_P = 2.18 \times 10^{-8}$ kg, then

$$\alpha_G = \frac{Gm_P^2}{\hbar c} = 1, \tag{16.15}$$

since $m_P^2 \equiv \frac{\hbar c}{G}$.  In this case the gravitational coupling is 137 times stronger than electrostatic coupling in the hydrogen atom, and would be stonger for larger masses.

A lesson from this is that even though the gravitational force (or coupling) is thought of as being the weakest force, it can be the strongest of them all since it does not saturate. It depends on how much mass is involved.

## 17 Holeum Instability

Chavda and Chavda (2002, p. 2928) propose that black holes and holeum are created, "When the temperature of the big bang universe is much greater than $T_b = mc^2/k_B$, where m is the mass of a black hole and $k_B$ [ k here] is the Boltzmann constant...." Let us examine whether the binding energy is great enough to hold holeum together in this high temperature regime. The binding energy between the masses m and m is given by j = 1 in (Rabinowitz, 2003) eq.(4.8) for 3-space, n = 3. In order for the binding energy given by eq. (4.8) to be large enough to hold the holeum atom together for high energy collisions in this regime, it is necessary that

$$E_n = \left[\frac{(n-4)}{(n-2)}\right]\left[\frac{G_n Mm \Gamma\left(\frac{n}{2}\right)}{\pi^{(n-2)/2}}\right]\left[\frac{\left[2G_n Mm^2 \Gamma\left(\frac{n}{2}\right)\right]}{(j\hbar)^{2(n-2)/(4-n)} \pi^{(n-2)^2/2(4-n)}}\right]^{(n-2)/(4-n)}, \tag{17.1}$$

$$E_{binding} = |E_{j=1}| = \frac{G^2 m^5}{4\hbar^2} \geq kT \gg kT_b = mc^2 \quad \text{for } n = 3,$$

where $kT \gg kT_b = mc^2$ is given by (Chavda,Chavda 2002), as quoted above. Equation (17.1) implies that

$$m \gg \sqrt{2}\left(\frac{\hbar c}{G}\right)^{1/2} = \sqrt{2}\, m_{Planck}. \qquad (17.2)$$

Equation (17.2) says that masses $\gg$ the Planck mass are needed for holeum to be stable in this high temperature regime. This is incompatible with their position (p. 2932) that they are dealing with black holes less than the Planck mass, "In this paper, we consider black holes in the mass range $10^3$ GeV/$c^2$ to $10^{15}$ GeV/$c^2$." This limits the black hole masses from $10^{-24}$ kg to $10^{-12}$ kg, compromising the stability of holeum by tens of orders of magnitude. Both for stability and to circumvent the need for a theory of quantum gravity, masses $\geq 2 \times 10^{-8}$ kg = $m_{Planck}$ are required. But this brings in problems of too small an orbital radius as shown next.

A mass $fm_{Planck} = f(\hbar c/G)^{1/2}$, where f is a pure number can be substituted into eq. (16.2) for n = 3 and j = 1 to ascertain the orbital radius, i.e. the separation of the two black holes for the ground state of holeum.

$$r_{j=1} = \frac{2\hbar^2}{G[fm_P]^3} = \frac{2\hbar^2}{G\left[f\left(\frac{2\hbar c}{G}\right)^{1/2}\right]^3} = \frac{2}{f^3 c}\left[\frac{\hbar G}{2c}\right]^{1/2}. \qquad (17.3)$$

Let us compare this radius with the black hole $R_H = 2Gm/c^2$ for $m = f(\hbar c/G)^{1/2}$,

$$\frac{r}{R_H} = \frac{2}{f^3 c}\left[\frac{\hbar G}{2c}\right]^{1/2}\left[\frac{c^2}{2Gf(2\hbar c/G)^{1/2}}\right] = \frac{1}{f^4}. \qquad (17.4)$$

For a stable orbit, f = $\sqrt{2}$, as determined by eq. (17.2). This implies that r = $R_H/4$. This is inconsistent with their use of Newtonian gravity (NG) which requires r > 2 $R_H$ just to avoid collision between the orbiting black holes. In NG, for equal black hole masses, each LBH orbits at a radius r/2 about the center of mass of the atom. For r > $10R_H$, NG requires f < $1/10^{1/4}$ = 0.56, but then the masses are each 0.56 $M_{Planck}$, requiring quantum gravity. For some cases they have $2R_H < 2r < 10R_H$, which is still not adequate.

Higher dimensional atoms will not alleviate this conundrum for the mass or the radius, as shown in (Rabinowitz, 2003). The way out of this problem is shown in Sec. 16 to have the mass M be a little black hole which is massive, yet with $R_H \ll r$, and an ordinary orbiting mass m $\ll M_{LBH}$. Note that higher dimensional bodies in n-space could be stable if they are like nucleons bound by short-range fields such as the Yukawa potential. The finding that orbiting bodies in atoms or planets cannot be bound for higher than 3-space applies only to long-range fields like the gravitational and electrostatic fields. For the sake of completeness, we next examine some of the limitations imposed by higher dimensional space.

## 18 Limitations Imposed by Higher Dimensional Space

### 18.1 Introduction

Higher dimensions clearly give more degrees of freedom, but they also impose unexpected limitations and restrictions. Some theories conjecture that accelerated expansion of the universe and dark energy can be understood in terms of higher dimensional space. This author has earlier shown that angular momentum cannot be quantized in the usual manner (radius adjusting itself so that an an integral number of wavelengths can be accommodated in an orbit) in 4-space because the radius cancels out of the dynamic equations (Rabinowitz, 2003). This leads to novel opportunities such as the quantization of mass. Now a distinct derivation will be presented that stable gravitational or electrostatic orbits are not possible for spatial dimensions n ≥ 4. Atoms and planets cannot be bound in higher dimensions. String theory may be impacted since the unfurled higher dimensions of string theory will not permit the existence of stable atoms. String theory with compacted dimensions may even be incompatible with accelerated expansion of the universe. The search for deviations from $1/r^2$ of the gravitational force at sub-millimeter distances may also be impacted. Nevertheless, I do not view string theory as an exercise in metaphysics. Finkelstein (1997) said that compacted dimensions "could have physical meaning... ." I agree with this, and think that taking string theory seriously also involves questioning it strenuously, as well as considering the limitations imposed by higher dimensional space.

A reasonable question to ask of any compacted higher dimensional theory is: What would the universe be like if the compacted dimensions unfurl to macroscopic dimensions? This paper attempts to answer this question by a novel demonstration that gravitational and electrostatic orbits are not stable for spatial dimensions n ≥ 4. It is thus shown that in higher dimensional space atoms cannot exist and that planetary motion is not possible. For convenience some previously derived results will be used (Rabinowitz, 1990 and 2003).

**18.2  No energetically bound circular orbits for n > 3 in n-space**

We will first examine circular orbits for n ≥ 3, and then build from these results to reach the same conclusion for general orbits. Gravitational and electrostatic long-range attractive forces can be expressed in n-space  n = 3, 4, 5, ... , as

$$F_n = \frac{-K_n}{r_n^{n-1}}. \tag{18.1}$$

For the gravitational force (Rabinowitz, 2001 a, b, and 2003)

$$K_{Gn} = \frac{2\pi G_n M m \Gamma(n/2)}{\pi^{n/2}}, \tag{18.2}$$

where we will consider the orbiting mass m << M. For the electrostatic force

$$K_{En} = \frac{2\pi R_{En} Q q \Gamma(n/2)}{4\pi\varepsilon\pi^{n/2}}, \tag{18.3}$$

where a body of mass m with negative charge q orbits around a positive charge Q. $R_{En}$ is a model dependent factor that relates the electrical force in n-space to the electrical force in 3-space, and $\varepsilon$ is the permittivity of free space.

Equating $F_n$ to the centripetal force, yields the kinetic energy:

$$\frac{-K_n}{r_n^{n-1}} = \frac{-mv_n^2}{r_n} \Rightarrow \tfrac{1}{2}mv_n^2 = \frac{K_n}{2r_n^{n-2}}. \tag{18.4}$$

The potential energy is

$$V_n = -\int \vec{F}_n \bullet d\vec{r} = \frac{-K_n}{(n-2)r_n^{n-2}}. \tag{18.5}$$

The total energy is

$$\begin{aligned}E_n &= \tfrac{1}{2}mv_n^2 + V_n = \frac{K_n}{2r_n^{n-2}} + \frac{-K_n}{(n-2)r_n^{n-2}} \\ &= \left[\tfrac{1}{2} - \frac{1}{(n-2)}\right]\frac{K_n}{r_n^{n-2}} = \left[\frac{n-4}{n-2}\right]\frac{K_n}{2r_n^{n-2}}\end{aligned} \tag{18.6}$$

The total energy $E_n \geq 0$, for $n \geq 4$. This result applies no matter how strong the attractive force, both classically and quantum mechanically, as quantization will not change the sign of the co-factor $K_n / r_n^{n-2}$. For a bound orbit in 3-space, $E_3$ must be negative, i.e. < 0. It is a little more complicated in higher dimensions to determine if an orbit is bound. Let's look at this next.

## 18. 3 Non-circular orbits in higher dimensions

In higher dimensional space central force trajectories are generally neither circular, nor elliptical, as the orbits become non-closed curves. The derivation in this paper is general, and differs from the approach taken long ago by Ehrenfest (1917, 1920), notwithstanding some similarities. We next show that for $n \geq 4$, circular orbits must be at the peak of the effective potential energy curve V' (cf. Fig. 1) and hence are unstable. The general case can be put in the form of a one-dimensional radial problem in terms of the effective potential energy of the system,

$$V'_n = V_n + L^2 / 2mr_n^2. \tag{18.7}$$

where $V_n(r)$ is the potential energy of the system, and L is the conserved angular momentum.

The orbits are not bound if $E_n - V'_n(r_m) \geq 0$, where $r_m$ is the radius of the circular orbit at the maximum of $V'_n$ (cf. Fig. 1). If atoms could be formed in the region $0 < E_n < V'_n(r_m)$ they would be only metastable since the finite width of the potential energy barrier presented by $V'_n$ permits the orbiting body to tunnel out as illustrated in Fig. 1.

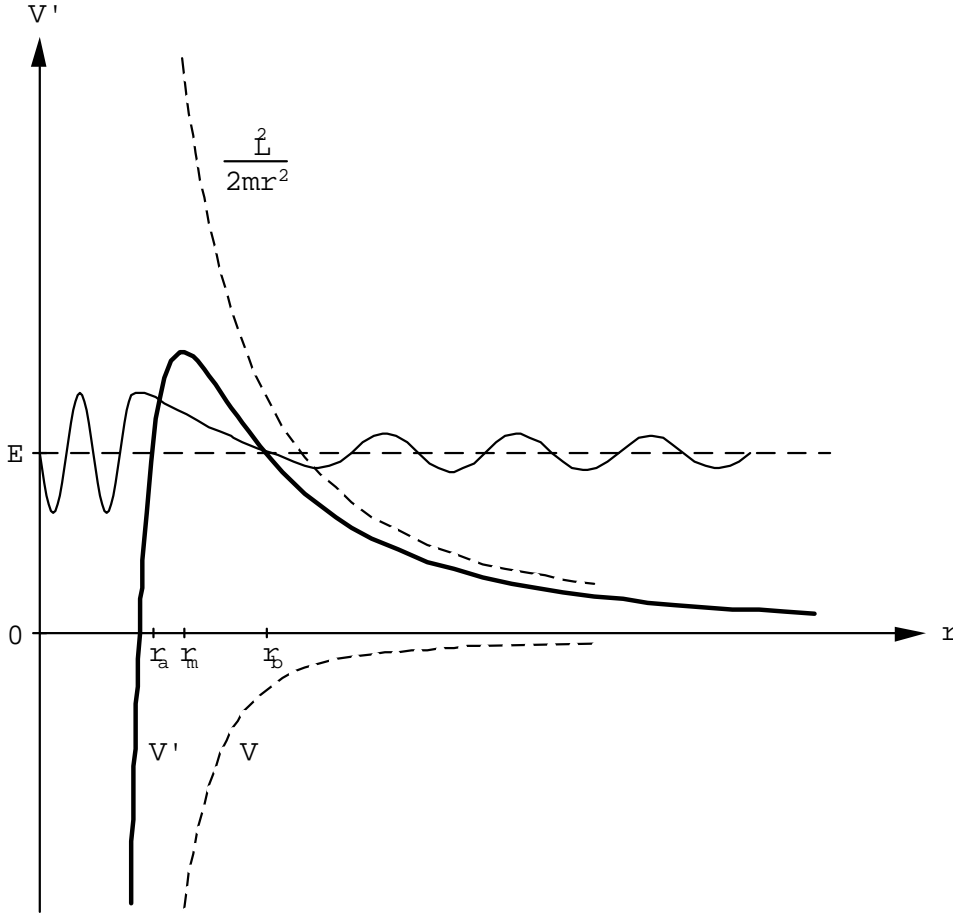

**Figure 1**. Effective potential energy in n-space with maxium value at $r_m$, showing wave function tunneling through the finite barrier of width $r_b - r_a$ at total energy $E > 0$.

The general equation of motion that includes radial motion is

$$F_n = \frac{-K_n}{r_n^{n-1}} = m\frac{d^2r}{dt^2} - \frac{mv_n^2}{r_n} = m\frac{d^2r}{dt^2} - \frac{L^2}{mr_n^3}. \tag{18.8}$$

Let us determine if there is an n that satisfies:

$$E_n(r_m) - V_n'(r_m) = E_n + \frac{K_n}{(n-2)r_m^{n-2}} - \frac{L^2}{2mr_m^2} \geq 0. \tag{18.9}$$

The maximum value of $V_n'$ occurs at $r_m$, and is obtained by setting $dV_n'/dr = 0$.

$$\frac{-K_n}{r_m^{n-1}} = \frac{-L^2}{mr_m^3} \Rightarrow r_m = \left[\frac{mK_n}{L^2}\right]^{1/(n-4)}. \tag{18.10}$$

This is the radius $r_m$ for a circular orbit at the maximum value of $V_n'$. This is an unstable orbit as any perturbation will change the nature of the orbit. Trajectories with $r > r_m$ are unbound both classically and quantum mechanically as can be seen from Fig. 1. Substituting for $E_n$ from eq. (18.6) into eq. (18.9),

$$\left[\frac{n-4}{n-2}\right]\frac{K_n}{2r_m^{n-2}} + \frac{K_n}{(n-2)r_m^{n-2}} - \frac{L^2}{2mr_m^2} \geq 0. \tag{18.11}$$

Combining the first two terms, and substituting eq.(18.10) into eq. (18.11):

$$\frac{K_n}{r_m^{n-2}} \geq \frac{L^2}{2m}r_m^{-2} \Rightarrow 1 \geq \frac{L^2}{2mK_n}\left(r_m^{n-4}\right) = \frac{L^2}{2mK_n}\left(\left[\frac{mK_n}{L^2}\right]^{1/(n-4)}\right)^{n-4} = 1. \tag{18.12}$$

Equation (18.12) implies that the circular orbit at $r = r_m$ is at the highest energy state, and thus

$$E_n(r_m) = V_n'(r_m) > E_n(r_n) \tag{18.13}$$

Let's first look at $E_n$ non-relativistically by means of the uncertainty principle with $p \sim \Delta p \sim \hbar/2\Delta x$, and $r \sim \Delta x$:

$$E_n \sim \frac{(\Delta p)^2}{2m} - \frac{K_n}{(n-2)(\Delta x)^{n-2}} = \frac{\hbar^2}{8mr^2} - \frac{K_n}{(n-2)r^{n-2}}. \tag{18.14}$$

We can conclude from eq. (18.14) that for $n \geq 4$, and r small enough to make $E_n < 0$, the orbiting body would spiral in to $r = 0$ both quantum mechanically and classically since then the negative potential energy term dominates in eq. (18.14). For large kinetic energies, this needs to be checked relativistically, which will be done next.

Let us look at $E_n$ by means of the uncertainty principle with $r \sim \Delta x$, and the relativistic energy equation:

$$\begin{aligned}E_n &= \left[(pc)^2 + m_0^2c^4\right]^{1/2} + \frac{-K_n}{(n-2)r_n^{n-2}} \\ &\sim \left[\left(\frac{\hbar}{2r_n}\right)^2 c^2 + m_0^2c^4\right]^{1/2} + \frac{-K_n}{(n-2)r_n^{n-2}}\end{aligned} \tag{18.15}$$

eq. (18.15) indicates that for $n \geq 4$ and r small enough to make $E_n < 0$, the orbiting body would spiral in to $r = 0$ both quantum mechanically and classically since then the negative potential energy term dominates in eq. (18.15).

Therefore quantum mechanically for $n \geq 4$, orbits of any configuration are not bound. Classical orbits can exist in the region $0 < E_n < V_n'(r_m)$. However, since they would be subject to quantum tunneling as depicted in Fig. 1, classical orbits would only be metastable. For $n \geq 4$ and r small enough to make $E_n < 0$, the orbiting body would spiral in toward the center of force at $r = 0$ both quantum mechanically and classically since then the negative potential energy term dominates in eqs. (18.14) and (18.15).

## 18.4 Discussion

A framework combining hierarchy theory (Dirac, 1937 and 1938) and string theory was proposed by postulating the existence of 2 or more compact dimensions in addition to the standard 3 spatial dimensions that we commonly experience (Argyres, 1998). In this view, gravity is strong on a scale with higher-dimensional compacted space, and only manifests itself as being weak on a macroscopic 3-dimensional scale.

One prediction (Arkani-Hamed, 1998) is that if there are only 2 compacted dimensions of compacted radius $r_c \sim 10^{-2}$ cm, it should be possible to detect a deviation of the Newtonian $1/r^2$ force law at this scale. It has been shown in this paper that for $r_c \sim 10^{-8}$ cm, common electrostatically bound atoms would not be stable, so a deviation of the Newtonian $1/r^2$ force law may not occur above this scale. To say that the electrostatic force is stuck on the brane and makes atoms stable seems only definitional.

Most variations of string theory only modify gravity at short distances. Since they leave gravity unchanged on a large scale, they do not even address the question of cosmic expansion. Standard string theory may even be discrepant with respect to accelerated cosmic expansion. Why are 6 to 7 of the dimensions tightly curled and our three unfurled? Is extra energy needed to stabilize the tautly wound dimensions? Or are they like abstract deBroglie waves wound around atomic orbits, that seem not to care whether they follow a line, a circle, an ellipse, etc.

It has been shown that neither gravitational nor electrostatic quantized orbits are stable for spatial dimensions $n \geq 4$ no matter how strong the attractive force. Even though classical orbits can exist in the region $0 < E_n < V_n'(r_m)$, they would only be metastable as they would be subject to tunneling through the effective potential energy barrier which has a finite width.

## 19  Black Holes and Entropy of the Universe

### 19.1  Bekenstein entropy

In 3-dimensional space Bekenstein (1972, 1974) found that the entropy of a black hole is $S_{bh} \propto kAc^3/G\hbar$. It is ironic that the prime critic of Bekenstein's 1972 conception was Stephen Hawking (1975), who embraced the concept three years later and found the constant of proportionality to be $1/4$ so that

$$S_{bh} = kAc^3/4G\hbar = \frac{kc^3}{4G\hbar} 4\pi R_H^2 = \frac{\pi k c^3}{G\hbar}\left[\frac{2GM}{c^2}\right]^2 = 4\pi k \frac{M^2}{\left(\frac{c\hbar}{G}\right)}$$

$$= 4\pi k \left[\frac{M}{M_{Pl}}\right]^2 = k \ln N_s$$
(19.1)

where A is its surface area, M is the mass of the black hole, $M_{Pl} \equiv (c\hbar/G)^{1/2} = 2.18 \times 10^{-8}$ kg is the Planck mass, $k = 1.38 \times 10^{-23}$ J/K, and $k \ln N_s$ is the standard Boltzmann statistical mechanical entropy of a system containing $N_s$ distinct states. It is not clear what distinct black hole states are being counted by $N_s$ in the expression $S_{bh} = k \ln N_s$. A further problem is that since entropy and temperature are statistical quantities dealing with many bodies, what does it mean to speak of them with respect to a black hole viewed as a single body, which is all that the Schwarzschild solution deals with. But these oversights seem not to have been a deterrent. The $4\pi$ in eq. (19.1) seems out of place and aesthetically incongruous with the statistical entropy $k \ln N_s$, so that it is tempting to leave it out, and I haven't always resisted this temptation in the past.

It follows from Bekenstein's formulation that the entropy of black holes is tremendously greater than the entropy of ordinary bodies of the same mass. For example, our sun of mass $2 \times 10^{30}$ kg (~ $10^{57}$ nucleons) and radius ~ $10^9$ m (~$10^6$ miles) has entropy $S \approx 10^{57}$ k $\approx 10^{35}$ J/K, whereas a black hole of the same mass has entropy $S_{bh} \approx 10^{76}$ k $\approx 10^{53}$ J/K, $10^{18}$ times higher with a radius of only ~ $10^3$ m (~1 mile). If the universe were 95% full of such black holes, there would be $10^{23}$ of them with a total entropy of $10^{99}$ k $\approx 10^{76}$ J/K. This represents an excess entropy of $10^{41}$ times that of our universe if it were filled with stars like our sun. Thus there is a colossally higher entropic probability that the big bang produced black holes dominantly over ordinary matter. This is a possible partial solution to the conundrum of why the early universe appears to have so little entropy, because most of its entropy is hidden in black holes. It appears likely that a large percentage of the mass of the universe is composed of little black holes according to my model since they do not evaporate away nearly as fast by gravitational tunneling radiation (GTR) as by Hawking radiation. Interference with nucleosynthesis is not an issue for GTR, as it would be for Hawking radiation which would dissociate big bang nucleosynthesis products, ruining the presently agreeable predictions of light element abundances. One may well expect LBH to be a major constituent of the remnants of the big bang, but can't be according to Hawking radiation, since they would have evaporated away long ago.

The precise entropy increase over that presently inferred depends on the distribution of LBH masses and that of ordinary matter. The LBH entropy is sensitive to the mass distribution as it depends on $M^2$ per LBH. For example if we consider that 95% of the universe was initially composed of $10^{20}$ kg LBH with radius of only $10^{-7}$ m, there would be ~ $10^{33}$ such LBH, each with entropy of $10^{33}$ J/K ($10^{56}$ k) with a total entropy of $10^{66}$ J/K ($10^{89}$ k). This is still impressively high.

**19.2 Entropy of the universe**

Generalizing Bekenstein's equation by using the n-space black hole horizon radius $R_{Hn}$ as given by eq. (29) in (Rabinowitz, 2001 b), the entropy of a black hole in n-space is

$$S_{bhn} \propto \frac{kc^3}{4G_n \hbar}[R_{Hn}]^2 = \frac{kc^3}{4G_n \hbar}\left[\frac{4\pi G_n M \Gamma(\frac{n}{2})}{(n-2)\pi^{n/2} c^2}\right]^{\frac{2}{n-2}}$$

$$\propto k\left[\frac{\Gamma(\frac{n}{2})}{(n-2)\pi^{n/2}}\right]^{\frac{2}{n-2}}\left[\frac{M}{M_{Pl}}\right]^{\frac{2}{n-2}}$$

(19.2)

It is noteworthy that the contribution to the entropy of the universe for $M \geq M_{Pl}$ increases for smaller mass black holes when the dimensionality of n-space is $\geq 4$.

If our universe were a black hole then its entropy would be

$$S_{bhn} \propto k \left[ \frac{\Gamma(\frac{n}{2})}{(n-2)\pi^{n/2}} \right]^{\frac{2}{n-2}} \left[ \frac{M_U}{M_{Pl}} \right]^{\frac{2}{n-2}} . \quad (19.3)$$

$$\xrightarrow{n=3} k \left[ \frac{M_U}{M_{Pl}} \right]^2 = k \left[ \frac{10^{53} kg}{10^{-8} kg} \right]^2 = 10^{122} k = 10^{99} J/K$$

If $G_n$ were known, it would be possible to determine n, the dimensionality that would maximize the entropy of the universe treated as if it were a black hole.

Penrose (1987) insightfully raised a crucial question, that in my opinion has not been answered to this day. He pointed out that because the big bang model requires an extremely small initial phase space volume, the initial entropy of the universe is exceedingly small. This presents a critical problem for inflation theory which has been so widely accepted in spite of also employing speeds greatly in access of the speed of light. Measurements of cosmic microwave background density fluctuations are not all in support of inflation. The conundrum of why the early universe appears to have so little entropy may only in part have a solution in that LBH would give it a larger entropy. Their strongly repulsive radiation because of their primordial proximity would cause rapidly accelerated expansion.

Black hole "no-hair" theorems state that black holes can be completely characterized by a few variables such as mass, angular momentum, electric charge, and magnetic charge (monopoles). [10] But is this true at all scales? Perhaps the minute dimpling of a black hole's surface on the scale of the Planck area by gravitational perturbations due to external matter of the rest of the universe can contribute significantly to a black hole's entropy. For completeness, if the universe were a black hole, one should also take into consideration the entropy inside it as well as the entropy associated with its horizon area. In 3-dimensions for a total mass M composed of N smaller black hole masses, the internal entropy

$$S_{3\,int} = Nk \left[ \frac{M/N}{M_{Pl}} \right]^2 = \frac{k}{N} \left[ \frac{M}{M_{Pl}} \right]^2 \quad (19.4)$$

decreases substantially as N gets large.

## 19. 3  Contrasting views of black hole entropy

Since there is not the least experimental confirmation of what the entropy of a black hole is, it is only fair to look at some alternate views. Some argue that Bekenstein's expression must be incorrect since it is not an extensive quantity proportional to the sum of individual entropies as is ordinary entropy. Dunning-Davies (2003) criticizes the accepted black hole entropy with conviction, "... the entropy being an extensive quantity [$\propto M$], which the accepted black hole entropy expression (1) most certainly isn't."  Such an assertion should not be dismissed out of hand, and it will be considered in Sec. 20.

Winterberg (1994) derives an expression similar to that of Bekenstein, except that for him the black hole entropy is proportional to the (3/4) power of the area instead of being proportional to the area of the black hole.  Thus for a solar mass black hole instead of an entropy of $10^{76}$ k, Winterberg's entropy would be

$$S_{bh} \sim \left(10^{76}\right)^{3/4} k = 10^{57} k . \tag{19.5}$$

For a body with an ensemble of N particles, the statistical mechanical entropy $S \approx Nk$ (within a logarithmic factor).  So eq. (19.5) is approximately equal to the statistical mechanical entropy for the sun, which has $\sim 10^{57}$ baryons.

Perhaps the view most in opposition to orthodox black hole entropy is that of Hüseyin Yilmaz (1958, 1982), for whom there is none.  Jacob Bekenstein (1972, 1974) conceived of black hole entropy in response to the conundrum posed to him by John Wheeler:  *What happens to entropy that is put inside a black hole?  Does the entropy of the universe go down?*  For Yilmaz this is a non-sequitur since in his modification of Einstein's general relativity, there are no black holes -- at best there are only very dense grey holes.

## 20  In Search of Novel Black Hole Entropy

### 20. 1  A rough but simple heuristic derivation of black hole entropy

Black hole entropy should be compatible with thermodynamic entropy.  So our starting point will be the standard thermodynamic definition of entropy and my approach will be heuristic rather than rigorous at this time:

$$S_{bh} \equiv \int \frac{dQ}{T} \leq \frac{Mc^2}{T} . \tag{20.1}$$

It is clear that prior to the acceptance of black hole radiation, eq. (20.1) would be in trouble since a "truly black" black hole must have T = 0, or it would radiate.  Thus its entropy would be infinite for any finite mass, with the strange possibility that only an infinite mass black hole might have a finite entropy.  Bardeen, Carter, and Hawking (1973) considered their effective black hole temperature as not real.  For the real temperature, they averred "the effective temperature of a black hole is zero ... because the time dilation factor [red shift] tends to zero on the horizon." Particles that originate at or outside the  horizon of an isolated black hole must lose energy in escaping the gravitational potential of the black hole.   Temperature could be inferred for an LBH from the energy distribution of emitted particles.  In GTR there is relatively little red shift, as the particles tunnel through the barrier with undiminished energy.  What little red shift there is, starts far from the black hole horizon (Rabinowitz 1999 b, 2003).

The Hawking 1974 value for  temperature is a factor of 2 smaller than his 1975 value, and would change the factor of 1/4 he found in 1975 to 1/2 for the entropy expression. This is not critical, and we will use his 1975 expression

$$T = \left|\frac{\hbar c^3}{4\pi kG}\right|\frac{1}{M} = \left[2.46 \times 10^{23}\right]\left(\frac{1}{M}\right) °K \tag{20.2}$$

in eq. (20.1). The steps in the following heuristic derivation are my own, and not those taken by Bekenstein -- or for that matter anyone else to my knowledge.

$$\begin{aligned} S_{bh} &\leq Mc^2 \left[\frac{4\pi kGM}{\hbar c^3}\right] = \frac{4\pi kM^2}{(c\hbar/G)} = 4\pi k \left[\frac{M}{M_{Pl}}\right]^2 \\ &= 4\pi k \left(\frac{R_H c^2}{2G}\right)^2 \left(\frac{G}{c\hbar}\right) = \frac{kAc^3}{4G\hbar} \\ &= \frac{kA}{4\left(\frac{G\hbar}{c^3}\right)} = \frac{1}{4}\frac{A}{\ell_P^2} \end{aligned} \tag{20.3}$$

where A is the black hole area, $\ell_p$ is the Planck length = 1.62 x 10$^{-35}$ m, and $\ell_P^2$ is the Planck area, also dubbed the Wheeler area by Jacob Bekenstein (1972, 1974) in honor of John Archibald Wheeler. Since as Bekenstein's thesis advisor, Wheeler encouraged him in the derivation of the black hole entropy equation, which was a brilliant theoretical accomplishment. Bekenstein made statistical arguments [in the last form of eq. (20.3)] that the factor was $(\ell n2)/8\pi$ = 0.0276, rather than the 1/4 obtained by Hawking (1975), which increased the entropy by an order of magnitude without experimental consequence.

It may be troubling that the classical limit as $\hbar \to 0$ gives $S_{bh} \to \infty$ independent of the black hole size, much like $T \to 0$ gives $S_{bh} \to \infty$; and $\hbar \to 0$ makes $T \to 0$. Perhaps another way of intuiting this, is that a particle of mass M can only be confined within $R_H$ if its reduced Compton wavelength $\lambdabar_c = \hbar/mc \leq R_H$. Or equivalently, $M_{Pl} \to 0$ as $\hbar \to 0$. As $\hbar \to 0$, an infinite number of such particles could be contained in any size black hole. If this is not completely satisfying, this may motivate the reader for a better answer. Or better yet, a better expression for $S_{bh}$.

It may also be troubling that according to conventional wisdom, it would be easier for a black hole to swallow a very large and hugely massive object with $\lambdabar_c \ll R_H$, than a very small and light particle with $\lambdabar_c \gg R_H$. It would take a long time for the black hole to ingest the large massive object, and the uncertainty principle may prevent a very low mass particle from ever being devoured. The latter does seem anti-intuitive, but so do many things in the quantum realm.

As discussed in Sec. 19.2, one criticism of eq. (20.3) is that it is not an extensive quantity. That is, when one puts two black holes together to form one black hole, the total entropy of the new combined system is not the sum of the two separate entropies. It would be if $S_{bh}$ were proportional to M and hence $R_H$, rather than $M^2$ or $(R_H)^2$. Nor do I think it should be extensive, as one would expect the entropy to increase in such an irreversible process as the coalescence of two black holes (Bekenstein 1972). What is surprising, is that my heuristic derivation started from the standard thermodynamic definition of entropy, which is extensive, and yet the final result is not extensive because of the inverse relation between temperature and mass in black holes.

## 20.2 Extensive black hole Entropy

There are a lot of ways that one could write $S_{bh}$ to give the Bekenstein black hole entropy in the limit of large M, and to be approximately extensive for small M. An example of one of the simplest is:

$$S_{bh} \propto \frac{M}{M_p}\left(1+\frac{M}{M_p}\right) \xrightarrow{M\to\infty} \left(\frac{M}{M_p}\right)^2, \tag{20.4}$$

and

$$S_{bh} \propto \frac{M}{M_p}\left(1+\frac{M}{M_p}\right) \xrightarrow{M\to 0} \left(\frac{M}{M_p}\right) \tag{20.5}$$

Though such an expression may not be consistent with current theoretical thinking, it is perfectly compatible with experiments -- of which there are none.

## 20.3 Surface dimpling contribution to black hole entropy

Another possible criticism of the entropy equation is that it does not take into consideration dimpling of the black hole surface, like an orange skin around the horizon. In simple Newtonian terms, one may think of the black hole surface as an equipotential surface. Viewed this way, one would expect the surface to be rough due to the presence of matter throughout the universe, notwithstanding Wheeler's maxim that *black holes have no hair* (Rabinowitz 1999 b). The question is whether this dimpling makes a negligible or appreciable contribution to the area of the black hole, and hence to its entropy. The Planck area is a very small scale, so one might not expect much contribution from dimpling, but it is better to make at least a rough calculation rather than none at all.

### 20.3.1 *Planck size entropy dimples*

One tempting possibility that seems to follow from Bekenstein black hole entropy is that space is quantized, with the smallest length being the Planck length, $l_p$, or the smallest area being the Planck area. If we take this point of view, then the dimples that may contribute to black hole entropy can at the smallest be of radius $l_p/2$. Then the maximum number of Planck dimples (spheres) would be

$$N_{spheres} \approx \frac{4\pi R_H^2}{\pi(l_p/2)^2} = 16\left(\frac{R_H}{l_p}\right)^2, \tag{20.6}$$

where the packing fraction for monolayer coverage has been neglected, $\approx 0.91$ for hexagonal packing, $\approx 0.79$ for square packing, etc. Thus the total maximum entropy would be

$$S_{bh} \approx \frac{k}{4}\left[\frac{\left(4\pi R_H^2\right)}{l_p^2} + \frac{\left[16\left(\frac{R_H}{l_p}\right)^2 4\pi(l_p/2)^2\right]}{l_p^2}\right] = \frac{5k}{4l_p^2}\left(4\pi R_H^2\right). \quad (20.7)$$

This gives a maximum entropy increase of a factor of 5. This is significant.

### 20.3.2 *Entropy dimples ~ Compton wavelength of the universe*

There are physically meaningful lengths that are much smaller than the Planck length, $l_p$. For example the reduced Compton wavelength of the universe:

$$\lambda_{cu} = \frac{\hbar}{M_u c} = 2.2 \times 10^{-95} \text{ m} \ll l_p = 1.62 \times 10^{-35} \text{ m}. \quad (20.8)$$

Actually the reduced Compton wavelength $\lambda_c$ of any mass M greater than $M_{Pl}$ is less than $l_p$, since $l_p/\lambda_c = M/M_{Pl}$. So stars, planets, micro-meteorites, and most LBH have $\lambda_c < l_p$. For a monolayer coverage of such tiny dimples, we simply replace $l_p$ with $\lambda_c$ in the second term of eq. (20.7):

$$\frac{S_{bh}}{k} \approx \frac{1}{4}\left[\frac{\left(4\pi R_H^2\right)}{l_p^2} + \frac{\left[16\left(\frac{R_H}{\lambda_c}\right)^2 4\pi(\lambda_c/2)^2\right]}{\lambda_c^2}\right] = \frac{4\pi R_H^2}{4}\left[\frac{1}{l_p^2} + \frac{4}{\lambda_c^2}\right] \approx \frac{4\pi R_H^2}{\lambda_c^2}, \quad (20.9)$$

for $\lambda_c \ll l_p$. One point here is to illustrate that even though the Planck scale appears to be quite fundamental, it may not be all that fundamental. So much of what is going on in modern physics, such as String theory and Loop Quantum Gravity, is centered around the Planck scale. Yet there appears to be no convincing argument that it is more fundamental than other comparably small or smaller scales. Of course if the normalization factor remained $l_p^2$ for the second term -- instead of $\lambda_c^2$, then the result would be independent of $\lambda_c$ as given by eq. (20.7).

If we had used $\lambda_c = \lambda_{cu}$, eq. (20.9) would give a really big entropy! Such a enormously large entropy for the early universe would go a long way to answering the Penrose (1987) incisive observation that the entropy of the nascent universe is unexplainably too low. It also addresses his very important insight from another perspective, in saying that the entropy of the universe has not increased that much, because it was already much higher than expected. However, It would be open to the criticism that it is unfairly big because the number of dimples exceeds the number of particles in the universe.

So in the interest of fairness, let us consider an average dimple producing mass, $M_d$. Then the maximum number of dimples is limited to the ratio $M_u/M_d$. With this limitation, eq. (20.9) becomes

$$\frac{S_{bh}}{k} \approx \frac{1}{4}\left[\frac{\left(4\pi R_H^2\right)}{l_P^2} + \frac{\left[\left(\frac{M_u}{M_d}\right)4\pi(\lambdabar_{cu}/2)^2\right]}{\lambdabar_{cu}^2}\right] = \frac{1}{4}\left[\frac{\left(4\pi R_H^2\right)}{l_P^2} + \pi\left(\frac{M_u}{M_d}\right)\right]. \qquad (20.10)$$

Though smaller, this is still a formidably large entropy. It would be smaller if a mass smaller than $M_u$ were used and/or some $\lambdabar_c > \lambdabar_{cu}$. Note that here also the result is independent of $\lambdabar_c$.

All the above Newtonian type calculations would need to be done in the context of general relativity. But that would be gilding the lily. The purpose of these simple alternative entropy expressions is not to present them as being rigorous, or as a challenge to Bekenstein entropy, which has been solidly defended with theory and good thought experiments. However, they have a certain degree of unavoidable circularity to them since they need to be self-consistent within a given framework. These expressions are presented to show that as long as there is no experimental evidence, many options are possible. Even doing them using general relativity may not be the last word. General relativity becomes highly non-linear in the region of a black hole, and even Einstein was concerned about general relativity solutions in this context.

## 21 General Discussion

The momentum transfer of beamed GTR can either be the main mechanism or just contribute to the accelerated expansion of the universe. As the universe gets old and the LBH radiate away to nothing, GTR predicts that the accelerated expansion of the universe will slow down, and eventually the acceleration will cease. This is in contrast to the incessant exponential accelerated expansion of the universe that may be obtained from a positive cosmological constant, $\Lambda$, of Einstein's general relativity. A positive $\Lambda$ predicts that the universe will endlessly expand quicker and quicker to a bleak future. In this bleak future, in $\sim 10^{11}$ years, this unbridled expansion would result in a universe in which only $\sim 10^2$ galaxies not far from the earth would be visible to us. The further ones would experience a relativistic red shift that would move their light from the visible into frequencies too low to be seen.

Most approaches to the accelerated expansion of the universe have been to invoke the cosmological constant of EGR or modify EGR. Many modify our conventional concepts of space-time. This amounts to modifying gravity. My approach may be the only one that does not modify gravity. Which one is correct? I'd like to say that mine is, but it is too soon to tell. The more things a given theory can explain, the better.  Mine explains many things.  Some approaches do little more than address the accelerated expansion without tying into other things. Sometimes things are so tightly bound that if one things falls the whole unravels.  Fortunately that is not the case here, if the correlations are not borne out, my main idea of Gravitational Tunneling Radiation still holds.  If the many correlations are borne out, then it makes GTR very likely.  However, ultimately GTR needs to be experimentally judged on its own.

The accelerated expansion of the universe notwithstanding, the laws governing gravity are mature, are closely confirmed by experiment, and appear to be well understood.  On the other hand, the thermodynamics and in particular the entropy of general gravitational systems leaves much to be desired.  Their ability to self-organize seems to be at odds with the second law of thermodynamics in that the apparent entropy of the system decreases.  Perhaps if we look closely enough, or properly modify the entropic concept, the overall entropy will be found to increase.  Our view of entropy may have to be as pliable as our view of the conservation of energy.  There are also those who argue that the principle that entropy can only increase cannot be applied to the entire universe.  For them such an extension is an unwarranted extrapolation founded on neither fact based on experience, nor on indisputable principle.

Though black holes are far from being well understood, the general orthodox view is that black hole entropy is on a firm and well understood basis.  Nevertheless, we should be vigilant, lest we become too complacent to challenge common consensus.  Black hole entropy does not represent a quantum limitation to entropy, but rather a quantum extension of the entropy concept. Part of this Chapter has probed some of the implications of black hole entropy in 3-space and higher dimensions, its malleability, and the entropy of the universe in n-space.  Black hole entropy may prove to be as malleable as the conservation of energy has in the past so that whenever a violation threatened, a new kind of energy was legitimately found to save it.

Beamed black hole radiation in GTR is key to my model for accelerated expansion of the universe.  The history of black hole radiation is not unlike the history of many other important developments in science.  Though Hawking is generally credited with the monumental concept that black holes can radiate, the credit should really go to Yakov Zel'dovich.  Zel'dovich (1971) proposed the first model of radiation from a black hole, "The rotating body [black hole] produces spontaneous pair production [and] in the case when the body can absorb one of the particles, ... the other (anti)particle goes off to infinity and carries away energy and angular momentum."  This is quite similar to the model proposed by Hawking (1974) for radiation from non-rotating black holes.  This relevant prior work by Zel'dovich was not referenced by Hawking in his original papers (1974, 1975), though he did reference some other work by Zel'dovich.  Neither Hawking nor Zel'dovich radiation has been detected, nor could either contribute to accelerated expansion of the universe because they would be isotropic.

## 22 Conclusion

Dark matter/dark energy is one of the deepest and darkest enigmas of modern science. The discovery of the accelerated expansion of the universe raises the question: What is causing this universal speedup? And nobody knows what or why? Theorists were quick to coin the term "dark energy" in concert with the already existing conundrum of "dark matter." The two have been considered as separate entities where dark energy is the cause of the accelerated expansion of the universe; and dark matter is the glue that holds the universe, or at least galaxies together. In my model, both are essentially the same, or due to the same source i.e. little black holes (LBH). As 95% of the mass of the universe, LBH essentially hold the universe together gravitationally, and their directed radiation causes its accelerated expansion. My model is the only one in which matter-in-the-universe contributes to it's accelerated expansion. All the others are rather insensitive to the constituents of the universe. The conventional view is that 4 - 5 % is ordinary matter, 23 - 25 % is dark matter, and about 70 - 73 % is exotic unknown dark energy unrelated to matter.

Gravitational tunneling radiation of LBH can account for the expansion speedup of the universe. Since the LBH are so small, they are essentially smoothed over in attempts to detect them, and so seem to lack discreteness (graininess or clumpiness), giving the appearance of being pure energy. Einstein's cosmological constant and inflation have been considered as explanations. However a big cosmological constant makes the vacuum enormously more massive than is consistent with observation or quantum theory. Directed radiation from LBH is a possible explanation that does not have these problems. For the totality of LBH (95 % of the mass of the universe) this radiation is effective in accelerating LBH and the universe outward (Rabinowitz, 2002) -- yet is considerably less than Hawking radiation that would fry the universe. In turn the LBH drag ordinary matter along with them -- such as luminous galaxies and the stars in them by which we gauge the accelerated expansion. In our epoch, this radiation eventually escapes into the immense voids created by the expansion of the universe, and redshifts into invisibility.

The catastrophic ruin of an area of more than 1200 square miles in Tungus Siberia on June 30, 1908 is not understood to this day, The hypothesis that it was caused by a heavy little black hole was prematurely laid to rest by the use of an impact velocity that was likely two orders of magnitude too small. The disastrous energy discharge scales inversely as the square of the impact velocity. Thus they incorrectly concluded that several orders of magnitude too much energy ($10^{15}$ to $10^{17}$ J) would have been released, implying that it could not possibly have been done by a black hole. So an LBH may still be slowly gobbling up part of Siberia. However, my analysis lets us end on the more optimistic note that even if it were an LBH, it is much more likely that the LBH exited through the earth, and that such devastation by an LBH is not likely to happen again for over 100,000 years.

**Acknowledgment**

I would like to thank Phil Garcia and Steve Crow for helpful discussions, Misha Shmatov for the Dmitriev et al reference, and Professor Albert W. Overhauser for bringing the Chavda and Chavda paper to my attention.